\documentclass[iop]{emulateapj}
\pdfoutput=1

\usepackage{natbib}
\usepackage{tabularx}
\usepackage{float}
\usepackage{graphicx}
\usepackage{subfigure}
\usepackage{threeparttable}
\usepackage{amsmath}
\usepackage{amssymb}

\newcommand{\HI}{\ensuremath{\mbox{\rm \ion{H}{1}}}}
\newcommand{\HII}{\ensuremath{\mbox{\rm \ion{H}{2}}}}

\newcommand{\htwo}{\ensuremath{\mbox{H$_2$}}}
\newcommand{\msun}{\ensuremath{M_\odot}}

\newcommand{\sunits}{\mbox{\msun ~pc$^{-2}$}}

\newcommand{\kms}{\mbox{km~s$^{-1}$}}

\newcommand{\cm}{\mbox{cm$^{-2}$}}

\newcommand{\co}[1]{\mbox{$^{#1}$CO}}

\newcommand{\av}{\ensuremath{\mbox{$A_{\rm V}$}}}

\newcommand{\counits}{\mbox{K km s$^{-1}$}}

\renewcommand{\d}[1]{\ensuremath{\mbox{d}{#1}}}

\newcommand{\sighi}{\ensuremath{\mbox{$\Sigma_{\rm HI}$}}}
\newcommand{\sightwo}{\ensuremath{\mbox{$\Sigma_{\rm H_2}$}}}
\newcommand{\rhtwo}{\ensuremath{\mbox{$R_{\rm H_2}$}}}

\newcommand{\nhtwo}{\ensuremath{\mbox{$N(\rm H_2)$}}}
\newcommand{\nh}{\ensuremath{\mbox{$N_{\rm H}$}}}
\newcommand{\nhi}{\ensuremath{\mbox{$N_{0, \rm HI}$}}}
\newcommand{\gdr}{\ensuremath{\mbox{$\rm{GDR}_0$}}}

\shortauthors{Imara \& Burkhart}

\begin{document}

\title{The \HI~Probability Distribution Function and the Atomic-to-Molecular Transition in Molecular Clouds}

\author{Nia Imara \& Blakesley Burkhart}
\affil{Harvard-Smithsonian Center for Astrophysics,
    Cambridge, MA 02138}

\email{nimara@cfa.harvard.edu}

\begin{abstract}
We characterize the column density probability distributions functions (PDFs) of the atomic hydrogen gas, \HI, associated with seven Galactic molecular clouds (MCs).  We use 21 cm observations from the Leiden/Argentine/Bonn Galactic \HI~Survey to derive column density maps and PDFs.   We find that the peaks of the \HI~PDFs occur at column densities ranging from $\sim 1$--$2\times 10^{21}~\cm$ (equivalently, $\sim 0.5$--1 mag).  The PDFs are uniformly narrow, with a mean dispersion of $\sigma_{\rm HI}\approx 10^{20}~\cm$ ($\sim 0.1$ mag).  We also investigate the \HI-to-\htwo~transition towards the cloud complexes and estimate \HI~surface densities ranging from 7--16 \sunits~at the transition.  We propose that the \HI~PDF is a fitting tool for identifying the \HI-to-\htwo~transition column in Galactic MCs.

\end{abstract}


\section{Introduction}
Setting the stage for star formation in the local Universe is the transformation of neutral atomic hydrogen, \HI, into molecular hydrogen, since stars are observed to form in the coldest, densest regions within molecular clouds (MCs).  In galaxies, the atomic component of the interstellar medium (ISM) is expected to be organized in phases, the warm neutral medium (WNM) and the cold neutral medium (CNM), existing in near pressure equilibrium (e.g., Kulkarni \& Heiles 1987; Wolfire et al. 2003; McKee \& Ostriker 2007).  Molecular clouds are expected to form in the coldest gas which, due to its higher number density, can more efficiently shield from dissociating photons than warm gas.  Indeed, in the Milky Way and beyond, MCs are commonly observed to be associated with regions of high-column-density \HI~(e.g., Wannier et al. 1983, 1991; Elmegreen \& Elmegreen 1987; Chromey et al. 1989; Engargiola et al. 2003).  Both Galactic and extragalactic studies indicate that MCs are frequently surrounded by envelopes of \HI~having surface densities of $\sighi\sim 8-10~\sunits$, (e.g., Engargiola et al. 2003; Imara \& Blitz 2011; Lee et al. 2012).  Such observations suggest that \HI~is a necessary ingredient for MC formation. On the other hand, perhaps \HI~envelopes result from the photodissociation of molecular hydrogen, \htwo~(e.g., Allen et al. 2004; Heiner et al. 2011).  It is also possible that such \HI~envelopes result from a combination of MC formation and evolution processes.  Either way, these interpretations highlight the importance of understanding the function of atomic gas in MCs and stellar evolution.

The various roles \HI~may play in MC evolution have been studied from theoretical and observational standpoints.  First, some cold atomic clouds provide sufficient shielding against dissociation by the ultraviolet (UV) interstellar radiation field that molecular gas can form in their interiors.  The structure of photodissociation regions (PDRs), where the transition from atomic to molecular gas occurs, has been treated theoretically by authors including van Dishoeck \& Black (1986), Hollenbach \& Tielens (1997), Draine \& Bertoldi (1996), Browning et al. (2003), and Krumholz et al. (2008; 2009).  The latter investigated the PDR around a spherical cloud of solar metallicity, and estimated that a minimum surface density of $\sighi \sim 10~\sunits$ is required for \htwo~formation. Moreover, Goldbaum et al. (2011) noted that higher initial surface densities of \HI~envelopes may result in higher mass MCs.  Second, MCs may continue to accrete atomic gas for an extended period during and their formation (e.g., Chieze \& Pineau Des Forets 1989; Hennebelle \& Inutsuka 2006; V\'azquez-Semadeni et al. 2010; Goldbaum et al. 2011).  Authors including Goldbaum et al. (2011) and  V\'azquez-Semadeni et al. (2010) demonstrated that while stellar feedback and accretion of atomic gas compete in regulating the mass of MC, accretion can be the deciding factor in the total mass of MCs.  Also, accretion of atomic gas may shape the dynamics of MCs by maintaining turbulence (Chieze \& Pineau Des Forets 1989; Hennebelle \& Inutsuka 2006; Goldbaum et al. 2011).  A third role of atomic gas in MC evolution is the influence of the CNM and WNM on molecule and star formation, as Stanimirovi\'c et al. (2014) and Lee et al. (2015) demonstrated may be the case in the Perseus molecular cloud.

In this paper, we employ a useful tool for studying the \HI~enveloping MCs.  We derive the probability distribution functions (PDFs) of the atomic gas associated with a group of Galactic clouds, with the aim of determining what the properties of the \HI~PDFs tell us about the MC evolution as well as the conversion from atomic to molecular gas---the \HI-to-\htwo~transition.  In recent years, a great deal of attention has been paid to evaluating the PDFs derived from column density maps of MCs from dust extinction observations (e.g., Lombardi et al. 2006; 2008; 2011; Kainulainen et al. 2009; Alves et al. 2014) and from dust continuum emission observations (e.g., Lombardi et al. 2014; Schneider et al. 2013; 2015a; 2015b; Imara 2015).  A number of theoretical studies determine that the widths of the column density PDFs correlate with the level of turbulence in MCs (e.g., Padoan et al. 1997; Federrath et al. 2008; Klessen 2000), while the high-extinction regimes of the PDFs correlate with the amount of high-density, self-gravitating, and potentially star-forming gas in MCs (e.g., Collins et al. 2012; Federrath \& Klessen 2013; Clark \& Glover 2014; Ward et al. 2014).  Indeed, a number of observational studies have found that PDFs derived from dust measurements have a ubiquitous log-normal distribution over a narrow range of column densities and exhibit a characteristic power-law ``tail'' at high densities above visual magnitudes of $\av\sim 2$ (e.g., Kainulainen et al. 2009; Froebrich \& Rowles 2010; Lombardi et al. 2015).

However, a consensus on how to interpret the widths and \emph{low}-extinction regimes of the dust PDFs has not been reached. This is because the dust extinction or emission observations used to derive the column density map toward a given MC has an inherent lower limit that is defined by the noise level of the observations.  Therefore, the width of a PDF---which may or may not be a manifestation of the amount of turbulence in a cloud---depends on the lowest level contour used to determine that PDF (Lombardi et al. 2015).  

While a number of studies have been dedicated to understanding the column density PDFs of clouds from dust measurements, there are far fewer observational treatments of the PDFs of the atomic gas associated with MCs.  Recently, Burkhart et al. (2015) examined the \HI~PDF of the Perseus molecular cloud.  They found that the shape of the PDF is log-normal and much narrower than the PDF of the cloud derived from dust extinction measurements.  The column density at the peak of the \HI~PDF in Perseus is close to the \HI~column density at the \HI-to-\htwo~transition, $\sim 6$--8 \sunits, measured by Lee et al. (2012).  While this study leads to the tempting suggestion that the \HI~PDF may be used to determine the column density at the \HI-to-\htwo~transition, such a proposal would be more persuasive if similar evidence were found for a larger sample of clouds.

Here, for the first time, the \HI~column density PDFs for a large group of Galactic MCs are investigated in detail, side by side with the dust extinction PDFs of the clouds.  In \S 2, we summarize the observations used to conduct this study.  In \S 3, we present our methods and results on \HI~column density PDFs. We discuss implications of our results, including the \HI-to-\htwo~transition and its connection with the \HI~PDF, in \S 4.  Concluding remarks are presented in \S 5.

\section{Data}
To derive the dust column density PDFs of the molecular clouds, we use extinction maps generously provided by C. Lada, M. Lombardi, and collaborators, which they created with their near-infrared color excess method, NICEST (Lombardi 2009; Lombardi et al. 2011).   This technique, which uses \emph{JHK} photometric data from 2MASS (Skrutskie et al. 2006) to estimate the amount of extinction due to dust, results in maps in units of visual extinction, \av.  The angular resolutions of the extinction maps are $1\arcmin$ for Ophiuchus, $2\farcm 5$ for Perseus, $40\arcsec$ for California, and $1\farcm 5$ for Orion A, Orion B, MonR2, and the Rosette.  The $1-\sigma$ uncertainties of the maps depend on extinction and Galactic latitude and range from $\sim 0.1-0.5$ mag.  

We use 21-cm observations to determine the PDFs of the atomic hydrogen gas associated with the MCs. The \HI~data are obtained from the Leiden/Argentine/Bonn (LAB) Galactic \HI~Survey (Kalberla et al. 2005), which spans velocities from $-400$ \kms~to +400 \kms.  The survey has a half-power beam width of $0\fdg 6$, velocity resolution of $1.3~\kms$, and an rms noise level of 0.07 K. 

We use archival \co{12}~data from the 1.2 m telescope at the Center for Astrophysics (Dame et al. 2001) to help select the velocity range of \HI~emission for each cloud complex (see \S \ref{sec:maps}). The uniformly sampled data from this all-sky survey have an angular resolution of $8\farcm 4$ and a velocity resolution of 0.65 \kms.  The rms noise level of the final data cubes is $\sim 0.25$ K per 0.65 \kms~channel.  

\scalebox{.9}[1.0]{The adopted distances to the MCs are listed in Table \ref{table1}.}
\begin{table}\centering
\begin{center}
\begin{tabular}{lccc}
\multicolumn{4}{c}{Table 1: Observational Parameters of MCs.}\\
\tableline\tableline
Cloud    & Distance  & \HI~Velocities & $R_{\rm MC}$ \\ 
         &  (pc)     &  (\kms)        &             \\
\tableline
Ophiuchus (1)  & $119\pm 6$   &  -3 -- 10     & $5\fdg 2$  (10.8 pc)   \\
Perseus (2)    & $240\pm 13$  &  -5 -- 15     & $2 \fdg 8$ (11.7 pc)   \\
Orion A (3)    & $371\pm 10$  &  ~0 -- 18     & $2 \fdg 4$ (15.5 pc)   \\  
Orion B (3)    & $398\pm 12$  &  ~0 -- 18     & $3 \fdg 1$ (21.8 pc)   \\  
California (4) & $450\pm 23$  & -10 -- 10     & $3\fdg 6 $ (28.2 pc)   \\
MonR2 (5)      & $905 \pm 37$ &  ~3 -- 19     & $ 1\fdg 6$ (24.8 pc)   \\  
Rosette (6)    & $1330\pm 48$ &  ~4 -- 18     & $1 \fdg 6$ (35.9 pc)   \\ 
\tableline
\end{tabular}
\caption{Distances to clouds; \HI~velocity ranges; and the projected radii of clouds,$R_{\rm MC}$, from the NICEST maps.  \textbf{Distance References.} (1) Lombardi et al. (2008); (2) de Zeeuw et al. (1999); (3) Lombardi et al. (2010); (4) Lada et al. (2009);  (5) Racine (1968); (6) Lombardi et al. (2011)}\label{table1}
\end{center}
\end{table}

\section{Methods and Results}\label{sec:results}
\subsection{Column Density Maps}\label{sec:maps}
The column density of the \HI~associated with MCs depends on the chosen range of velocities and the spatial boundaries. For the latter, we use the NICEST extinction maps to define the perimeter of the various clouds:

\begin{tabular}{llc}
 & & \\
Ophiuchus: & $-10^\circ \le l \le 8^\circ $, &  $8^\circ \le b \le  22^\circ$, \\ 
Perseus:  & $155^\circ \le l \le  162^\circ$, &  $-27^\circ \le b \le -15^\circ$, \\ 
Orion A:  & $205^\circ \le l \le  217^\circ$, &  $-21^\circ \le b \le -18.5^\circ $, \\ 
Orion B:  & $201^\circ \le l \le 210^\circ $, &  $-17.5^\circ \le b \le -10^\circ $, \\ 
California:  & $155^\circ \le l \le  169^\circ$, &  $-15^\circ \le b \le  -5^\circ$, \\ 
MonR2:  & $211^\circ \le l \le 216^\circ $, &  $-15^\circ \le b \le -10^\circ $, \\ 
Rosette: & $204^\circ \le l \le 209.5^\circ $, &  $-4^\circ \le b \le -1^\circ $.\\
 & & 
\end{tabular}
\begin{figure*}[ht]
\includegraphics[width=\textwidth]{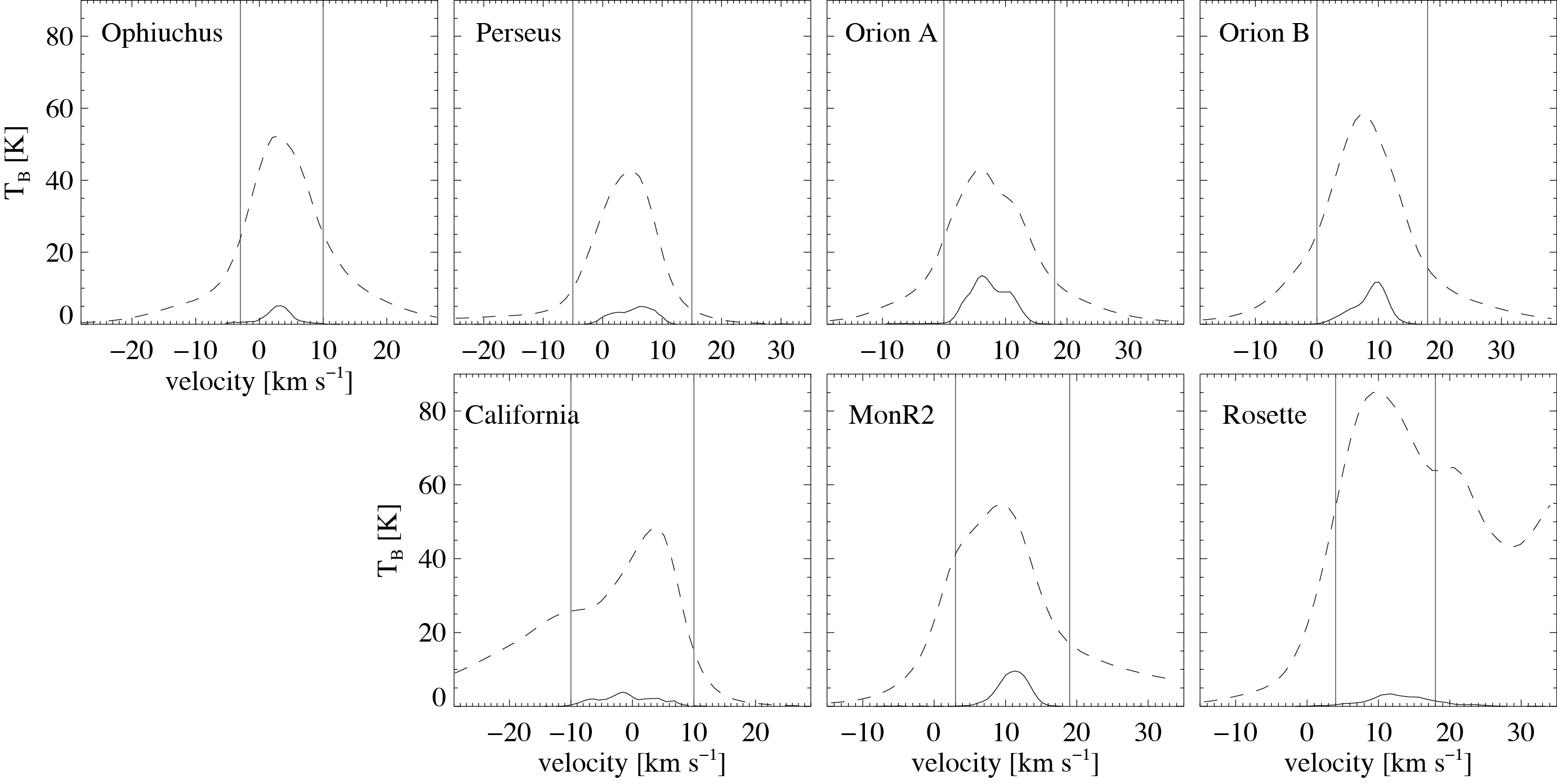}
\caption{Brightness temperature, $T_B$, versus velocity toward molecular clouds.  Dashed lines: average \HI~spectra.  Solid lines: average \co{12}~spectra times a factor of 10.  The vertical lines mark the velocity boundaries of the \HI~emission used in the analysis (see \S \ref{sec:maps}). \label{fig0}}
\end{figure*}

The boundaries contain most of the molecular emission associated with the clouds as determined by CO observations (see Dame et al. 2001).
To determine the spatial boundaries of the \emph{atomic gas envelopes} associated with MCs, we begin by convolving the NICEST maps to the resolution of the LAB data.  For each complex, we then select \HI~emission corresponding to all pixels in the NICEST map where $\av \ge 1$ mag. This criterion is motivated by observational studies demonstrating that dust extinction PDFs are typically difficult to characterize for $\av\lesssim 1$ mag (e.g., Lombardi et al. 2015).  Additionally, both observational and theoretical studies find that the \HI-to-\htwo~transition in local molecular clouds typically occurs around $\av=1$ mag (e.g., Krumholz et al. 2009; Lee et al. 2015).

To choose relevant velocity ranges of the \HI~associated with each cloud, following the basic methodology of Imara \& Blitz (2011), we begin by examining the CO spectra of the MCs from the Dame et al. (2001) 1.2 m CO Survey of the Galaxy.  In Figure \ref{fig0}, we plot the average \co{12}~spectrum through each MC (solid line) with the \HI~spectrum in the same direction (dashed line) overplotted.  Generally, the mean LSR velocities of the \HI~and of the CO emission are very close, to within a few \kms~of each other.  However, as demonstrated in Figure \ref{fig0}, the \HI~emission line is much broader than the CO line (e.g., Imara \& Blitz 2011).  Since we are concerned with investigating the atomic-to-molecular transition, we aim to avoid as much as possible including any \HI~that is unrelated to the MC complexes.  Thus, after a careful examination of the \HI~channel maps and spectra, we assume that \HI~emission having velocities within roughly a factor of 3 times the FWHM velocity dispersion of the \co{12} line is associated with a given molecular cloud.  This results in \HI~velocity ranges that are slightly narrower than those adopted by Imara \& Blitz (2011), (for Perseus, Orion A, Mon R2, and the Rosette), who selected \HI~velocities within $\sim 2$ times the velocity dispersion of the \HI~profile.  We note, in this study, we are concerned with restricting our analysis to \HI~emission corresponding as closely as possible to the molecular emission of the clouds, whereas the analysis of Imara \& Blitz (2011) did not require the same restrictions.  The defined velocity boundaries are shown as vertical lines in Figure \ref{fig0} and are listed in Table \ref{table1}, as well as the adopted distances to the clouds.  In \S \ref{sec:lowdensity}, we discuss the consequences of varying the kinematic and spatial boundaries.

We calculate the \HI~column density, $N(\HI)$, under the optically thin assumption, in which case the $N(\HI)$ is proportional to the \HI~brightness temperature, $T_B$.  For each cloud, the $N(\HI)$ is measured along each line of sight by integrating the 21-cm emission over the selected velocity range, according to,
\begin{equation}\label{eq:nhi}
N(\HI)=1.82\times 10^{18}\int_{v_{\rm min}} ^{v_{\rm max}} \frac{T_B} {\counits}~dv~\cm
\end{equation}
where $\d v$ is the channel velocity width. 

We relate the visual extinction, $\av$, of the NICEST maps to a total hydrogen column density via the gas-to-dust ratio conversion
\begin{equation}\label{eq:gdr}
\frac{\nh}{\av} = \frac{N(\HI)+2\nhtwo}{\av} = 1.87\times 10^{21}~\cm~{\rm mag}^{-1},
\end{equation}
of Bohlin et al. (1978), frequently used in other observational studies of extinction PDFs in the Galaxy (e.g., Kainulainen et al. 2009; Lombardi et al. 2011; Malinen et al. 2012; Alves et al. 2014).  In order to compare the NICEST PDFs and the \HI~column density PDFs in the following sections, we use this fiducial Galactic gas-to-dust ratio, \gdr, keeping in mind that variations have been found in the Galaxy.  Bohlin et al. (1978) pointed out that $N(\rm H)/\av$ can vary from \gdr~by as much as a factor of 1.5, while Kim \& Martin (1996) found that $N(\rm H)/\av$ can differ from the typical Galactic value by a factor of 3.  
\begin{figure}[!ht] 
\centering
\includegraphics[width=0.45\textwidth]{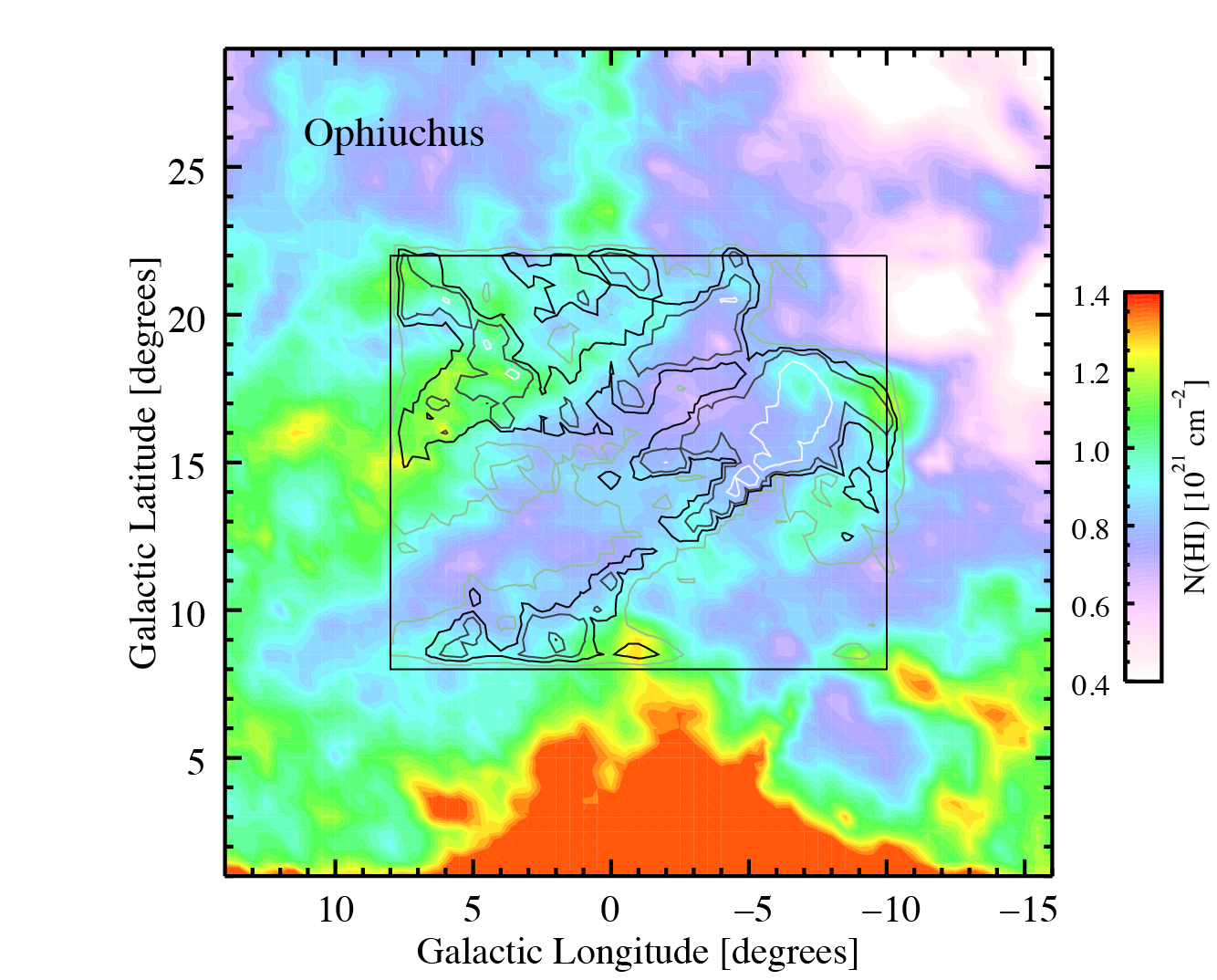}
\caption{Ophiuchus molecular cloud --- Column density maps of the atomic and molecular gas, convolved to the $0\fdg6$ resolution of the LAB data, in the vicinity of the Ophiuchus molecular cloud.  The dotted box delineates the perimeter of the NICEST-defined cloud, as given in Section \ref{sec:maps}, while the dimensions of the entire \HI~map are given by $R_A\sim 3R_{\rm MC}$ (see Table \ref{table1}).  The contour levels of the NICEST map, in units of $\av$, are 0.5, 1, 1.5, and 3 mag (light gray, black, gray, and white). The scale of the $N(\HI)$~map, in color, is indicated by the color bar. \label{fig1}}
\end{figure}

\begin{figure}
\centering
\includegraphics[width=0.45\textwidth]{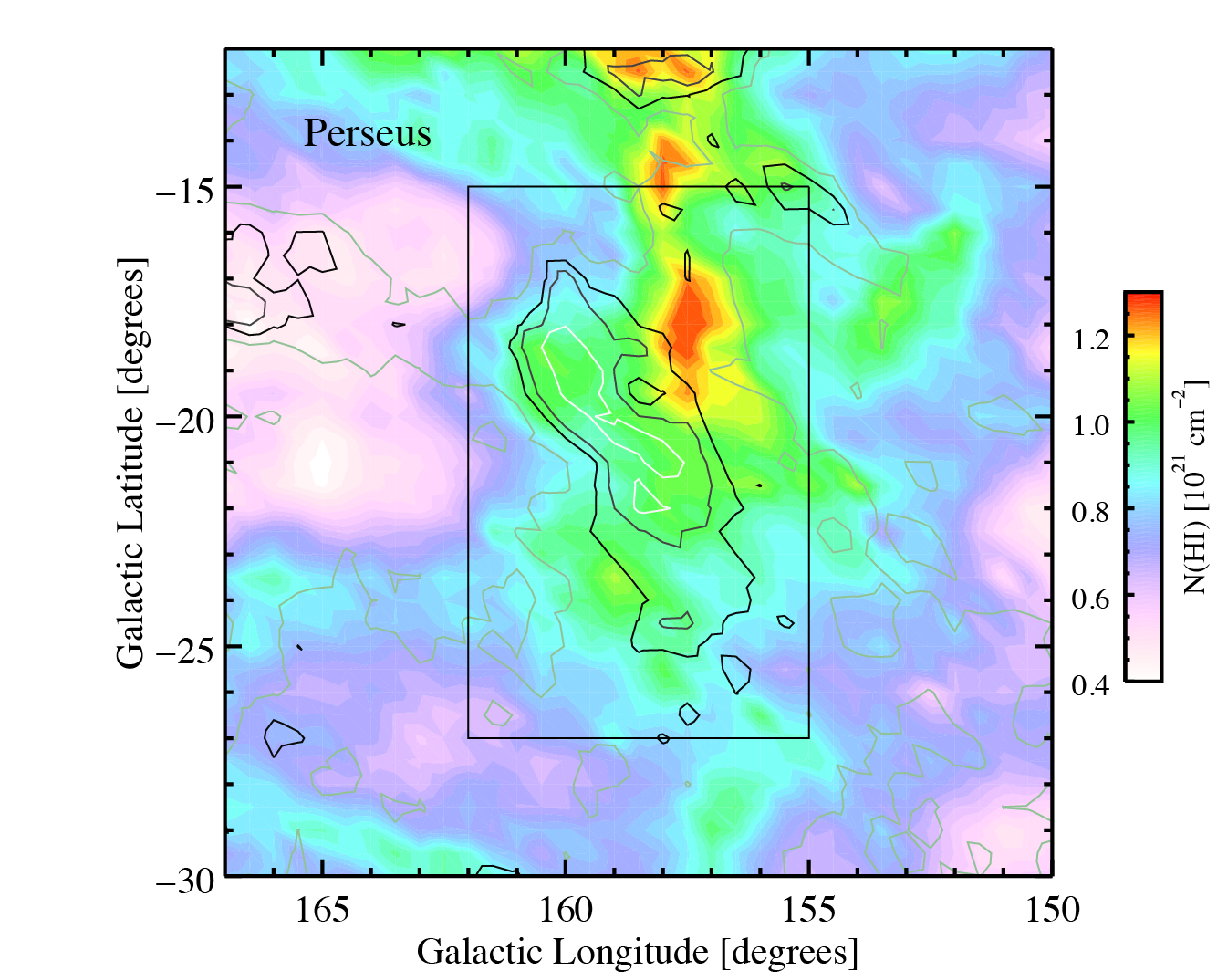}
\caption{Perseus molecular cloud --- Same as Figure \ref{fig1}.\label{fig2}}
\end{figure}

\begin{figure}[h]
\centering
\includegraphics[width=0.45\textwidth]{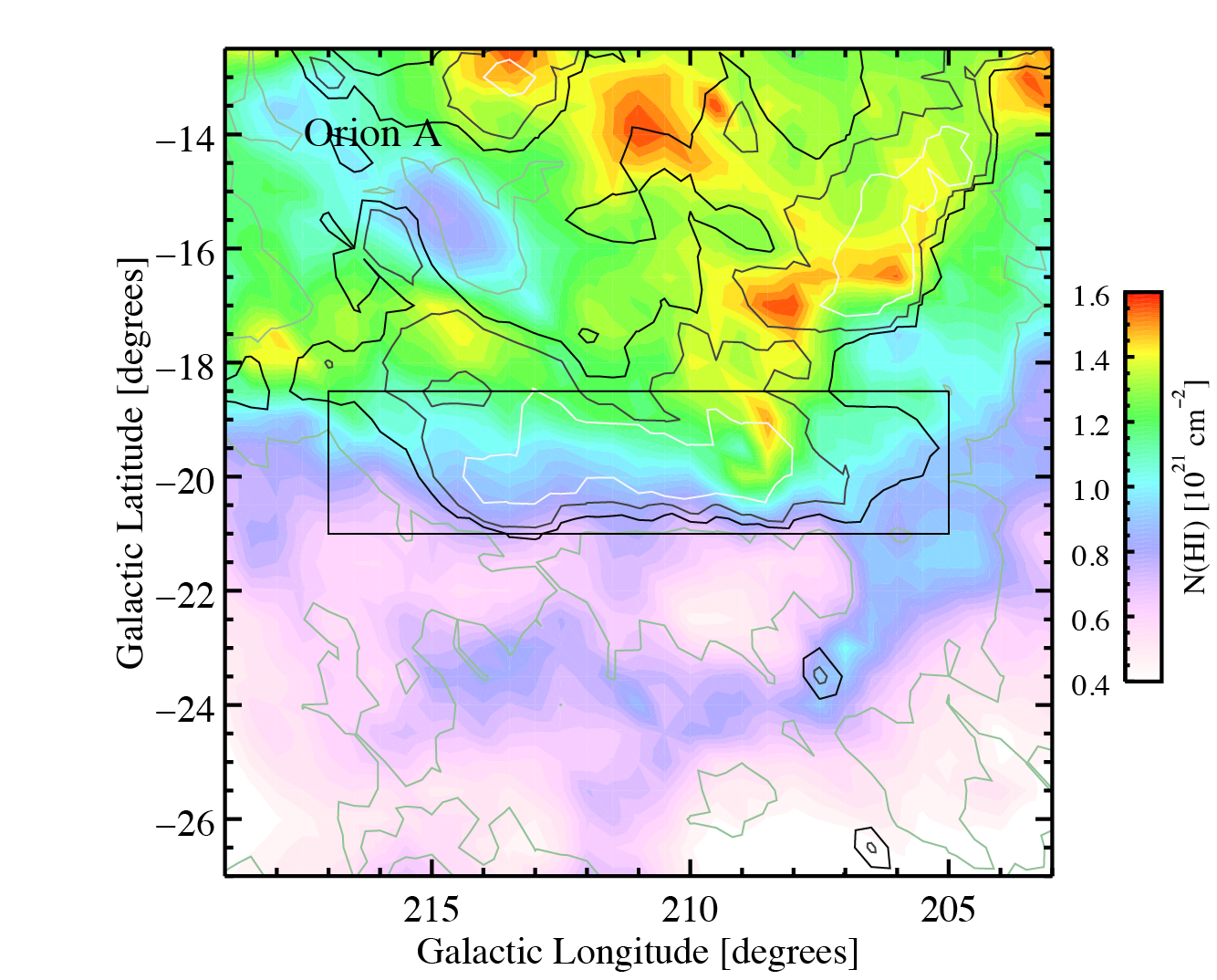}
\caption{Orion A molecular cloud --- Same as Figure \ref{fig1}. \label{fig3}}
\end{figure}

\begin{figure}[h]
\centering
\includegraphics[width=0.45\textwidth]{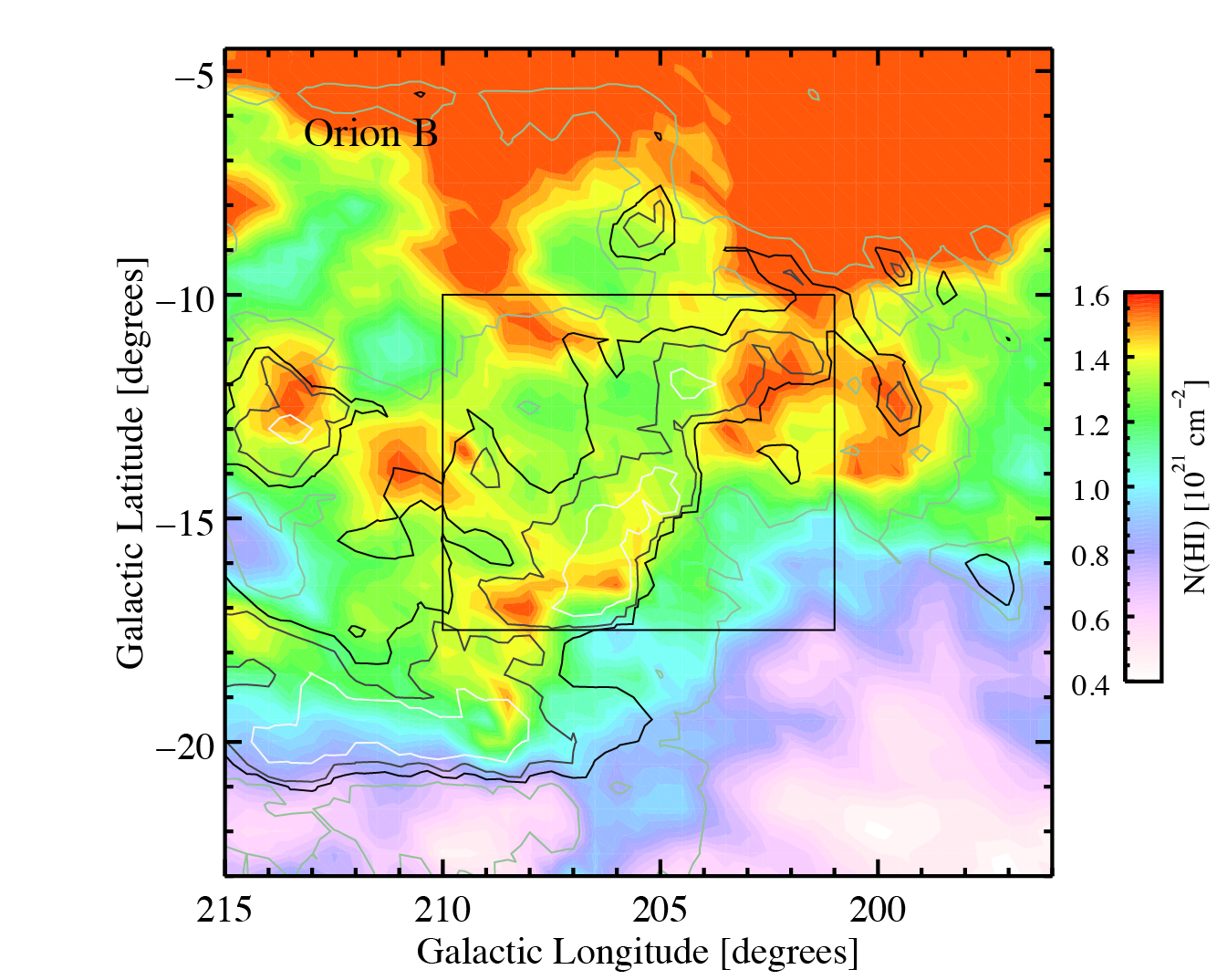}
\caption{Orion B molecular cloud --- Same as Figure \ref{fig1}. \label{fig4}}
\end{figure}

\begin{figure}[h]
\centering
 \includegraphics[width=0.45\textwidth]{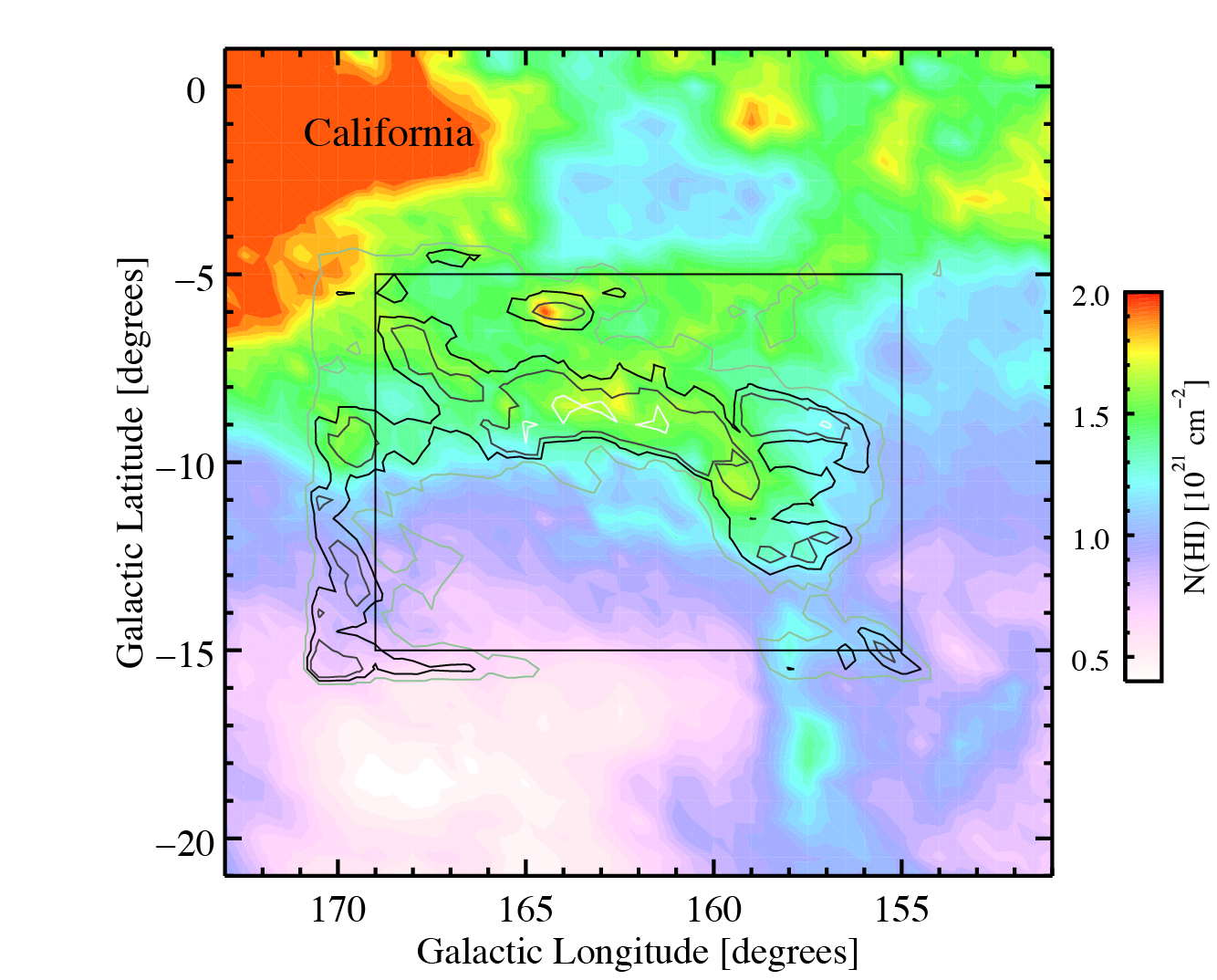}
\caption{California molecular cloud --- Same as Figure \ref{fig1}. \label{fig5}}
\end{figure}

\begin{figure}[h]
\centering
\includegraphics[width=0.45\textwidth]{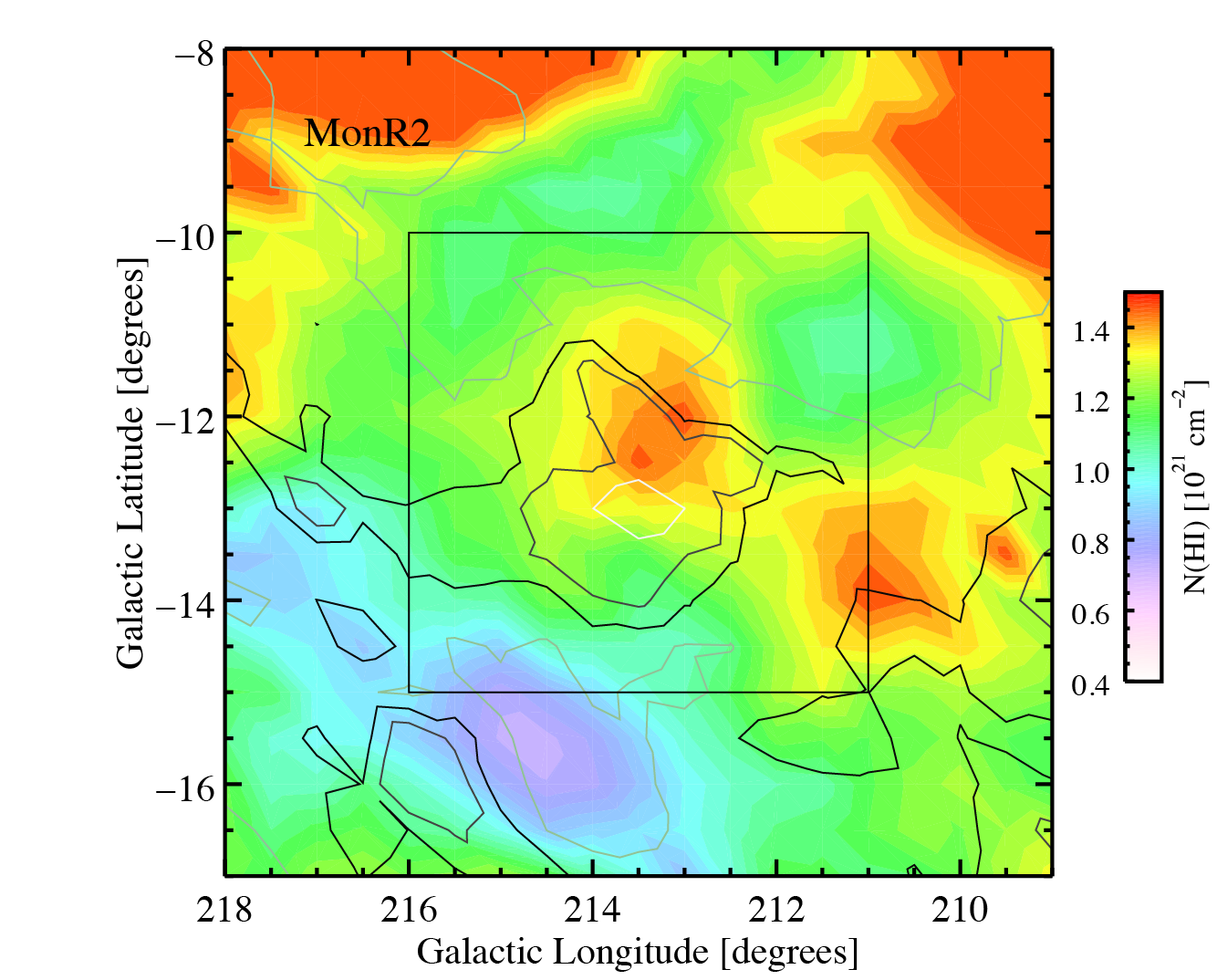}
\caption{MonR2 molecular cloud --- Same as Figure \ref{fig1}. \label{fig6}}
\end{figure}

\begin{figure}[h]
\centering
\includegraphics[width=0.45\textwidth]{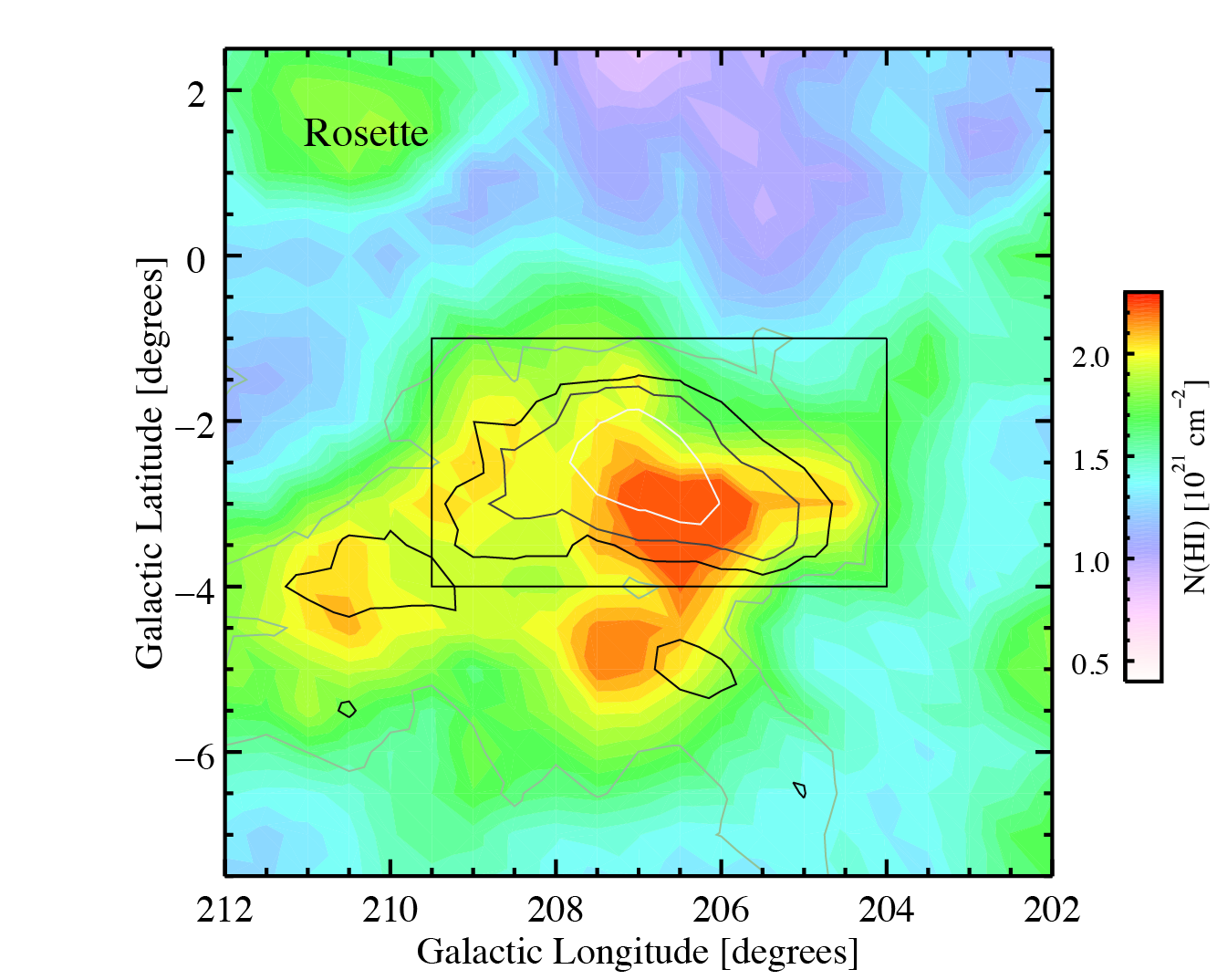}
\caption{Rosette molecular cloud ---  Same as Figure \ref{fig1}. \label{fig7}}
\end{figure}

Figures \ref{fig1} to \ref{fig7} display the resulting \HI~column density maps.  Overlaid in grayscale contours are the NICEST maps at the 0\fdg6 resolution of the \HI~maps.  To provide context about the environment surrounding the MCs, the sizes of the images in Figures \ref{fig1}--\ref{fig7} are larger than the defined spatial boundaries of the MCs.  The size of each image is roughly equal to the ``accumulation radius,'' $R_A\approx 3R_{\rm MC} $, the size of the region from which an MC must contract to have acquired its present mass. Following Imara \& Blitz (2011), the accumulation radius is defined by requiring that the mass contained within a cylinder of radius $R_A$, perpendicular to the Galactic plane, is equal to the present mass of the MC having radius $R_{\rm MC}$. The radii of the various MCs are estimated from the projected area, $A$, within the 1 mag contour of the NICEST maps, so that $R_{\rm MC}=\sqrt{A/\pi}$.  These values are listed in Table \ref{table1}.


\subsection{Column-density Probability Distributions}\label{sec:pdfs}
Figures \ref{fig8}--\ref{fig14} show the column-density PDFs for the seven cloud complexes in this study.  In each plot, the gray curve represents the PDF derived from dust extinction measurements using the NICEST method (Lombardi et al. 2009).  We note that in order to show the full dynamic range of the dust extinction measurements, the displayed NICEST histograms are derived from the maps at their original resolution.  We offer the high-resolution NICEST PDFs mainly as visual points of comparison with the \HI~PDFs, keeping in mind that the NICEST PDFs from the convolved, low-resolution maps wash out many of the salient features---especially at the low- and high-extinction regimes---we wish to highlight.

In Figures \ref{fig8}--\ref{fig14}, each red curve shows the distribution of \HI~column densities within the $\av\ge 1$ mag boundary of the given MC.  To provide context as to the behavior of the \HI~PDF in the larger region surrounding an MC, the pink curve shows the column density distribution of all \HI~gas within $R_A\sim 3R_{\rm MC}$ (i.e., the entire displayed field in Figures \ref{fig1}--\ref{fig7}). Unlike molecular clouds, which generally have well-defined, closed spatial boundaries in maps derived from molecular emission and dust extinction, diffuse atomic gas associated with MCs does not always have such distinct boundaries. 

For each cloud complex, the PDFs of the NICEST map and the \HI~within $\av\ge 1$ mag are each fitted with a log-normal distribution described by
\begin{equation}
p(\nh;\sigma)=\frac{1}{\sigma\sqrt{2\pi}}~\exp\left[{-\frac{(\ln~\nh - \ln~N_0)^2}{2\sigma^2} }\right],
\end{equation}
using the MPFITFUN least-squares fitting procedure from the Markwardt IDL Library (Markwardt 2009).  The column density at which the distribution peaks is $N_0$, and $\sigma$ is the dispersion.  In the remainder of this paper, the column density at which the \HI~PDF peaks and the dispersion of the \HI~PDF are referred to as $N_{0,\rm HI}$ and $\sigma_{\rm HI}$, respectively. The resulting fits to the \HI~PDFs are shown as red dashed lines in Figures \ref{fig8}--\ref{fig14} and presented with their 1-$\sigma$ uncertainties in Table \ref{table2}.

As discussed in detail in a number of studies, the PDFs derived from dust observations---in this case, dust extinction---seem to be described by a log-normal distribution over a limited range of column densities, while there is an excess over log-normal at high column densities above $\av\gtrsim 2$ mag (e.g., Kainulainen et al. 2009; Froebrich \& Rowles 2010; Lombardi et al. 2011; Lombardi et a1. 2014; Imara 2015; Schneider et al. 2015).  Frequently referred to as a ``tail,'' this excess at high column densities manifests as a power law in log-log space and  may arise from the gravitational collapse of gas related to star formation (e.g., Klessen 2000; Kritsuk et al. 2011; Federrath \& Klessen 2013; Clark \& Glover 2014; Girichidis et al. 2014; Ward et al. 2014; Schneider et al. 2013).  As Kainulainen et al. (2009) pointed out, PDFs from dust observations typically cannot be fit as a log-normal over their whole range.  Thus, for illustrative purposes only, we fit the NICEST PDFs up to $\av\le 2$ mag, and we show the resulting fits in Figures \ref{fig8}--\ref{fig14} with a dashed black line.

In Figures \ref{fig8}--\ref{fig14}, as a point of comparison, we also identify the location of $\av = 1$ mag for different values of the gas-to-dust ratio, $N_{\rm H}/\av$.  This is the extinction around which the \HI-to-\htwo~transition is predicted to occur by theorists (e.g., Krumholz et al. 2009).  Based on the result of Bohlin et al. (1978), who pointed out that the Galactic gas-to-dust ratio can vary from the fiducial value, $\gdr=1.87\times 10^{21}$ $\cm~{\rm mag}^{-1}$, by up to a factor of 1.5, we indicate $\nh/(1~{\rm mag})=\gdr$, $1.5\times \gdr$, and $\gdr/1.5$ with vertical lines in Figures \ref{fig8}--\ref{fig14}.

The \HI~PDFs are the focus of the remaining discussion. In \S 4.1, we discuss what the PDFs reveal about high-column-density \HI.  In \S 4.2, we discuss low-column-density gas, and we examine how the spatial and kinematic boundaries potentially influence the shape of the PDFs.  The \HI-to-\htwo~transition is the subject of \S4.3.  Finally, in \S4.4, we consider whether the dispersion of the \HI~PDF traces the sonic Mach number.

\section{Implications and Discussion}\label{sec:discussion}
Figures \ref{fig8}--\ref{fig14} show that the \HI~PDFs towards our sample of MCs (red curves) tend to have narrow, log-normal forms.  The pink curves in these figures show the PDFs of all the atomic gas centered within $\sim 3R_{\rm MC}$ of the cloud complex, and the gray curves are the PDFs from the NICEST extinction maps. In the following, we describe the properties of the PDFs in detail, we consider the various assumptions that potentially bias our results, and we discuss implications for the \HI-to-\htwo~transition.

\subsection{High-column-density \HI}
\begin{figure}[t]
\centering
\includegraphics[width=0.45\textwidth]{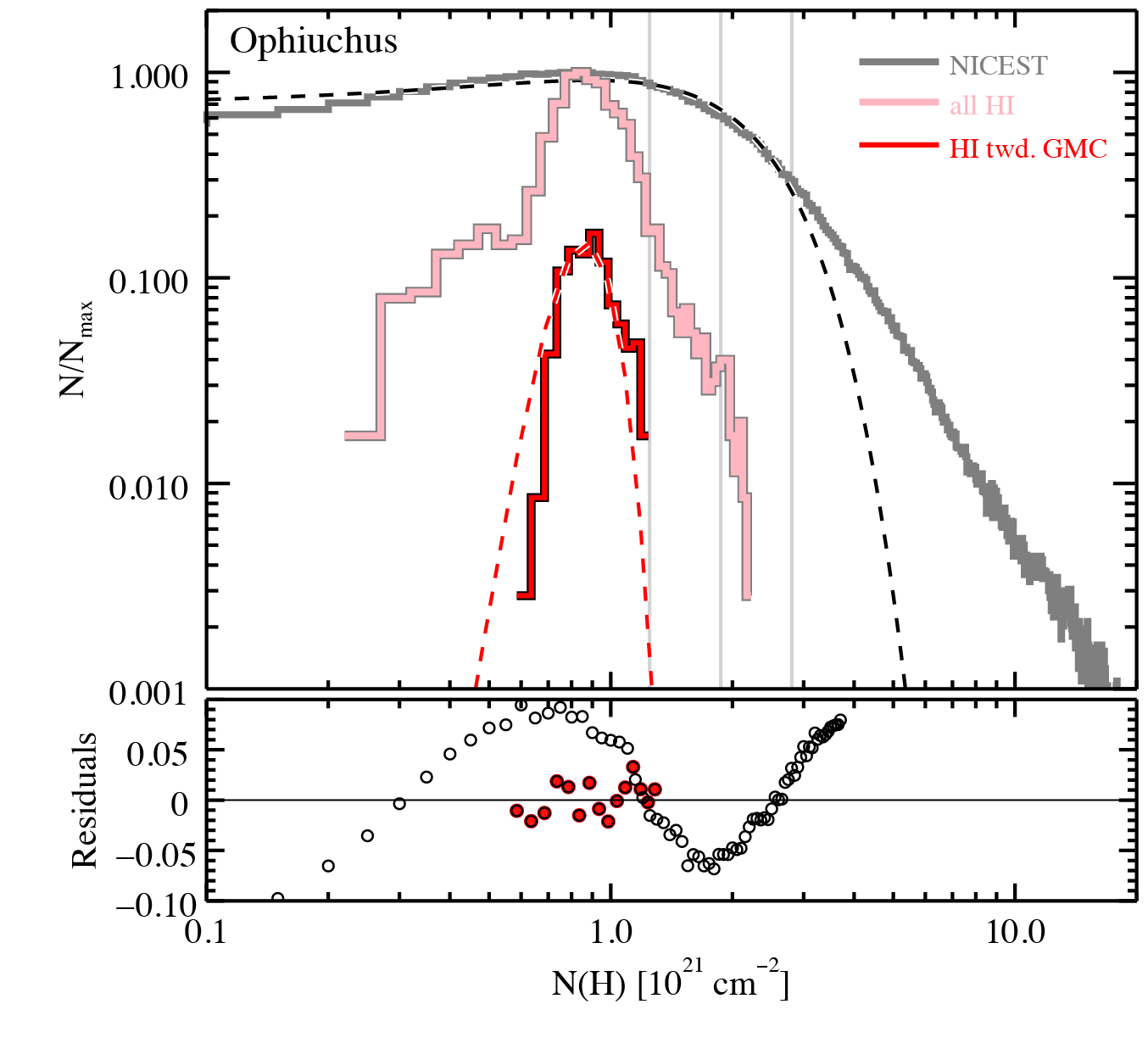}
\caption{PDFs of \HI~and \htwo~column densities in the Ophiuchus molecular cloud, with a bin size of $0.05\times 10^{21}~\cm$.  The gray curve represents the PDF obtained from the NICEST extinction map and is normalized to the bin containing the maximum number of pixels.  The red curve represents the PDF for \HI~emission~within the $\av \ge 1$ mag boundary of the NICEST extinction map and within the velocity range listed in Table \ref{table1}. The black and red dashed lines are Gaussian fits to the NICEST and \HI~PDFs, respectively.  The former is fit up to $\av\le 2$ mag, as specified in the text. The residuals to the fits are shown as white circles (NICEST) and red circles (\HI). The pink curve is the PDF for \HI~within $\sim 3R_{\rm MC}$ of the MC (i.e., corresponding to the entire \HI~field displayed in Figure \ref{fig1}).  The red curve is normalized to the maximum number of counts in the pink curve.  The vertical gray lines indicate the theoretical \HI-to-\htwo~transition at $\av = 1$ mag for three gas-to-dust ratios: the fiducial Galactic value \gdr~(solid line), $1.5\times\gdr$, and $\gdr/1.5$ (dotted lines).    \label{fig8}}
\end{figure}

\begin{figure}
\centering
\includegraphics[width=0.45\textwidth]{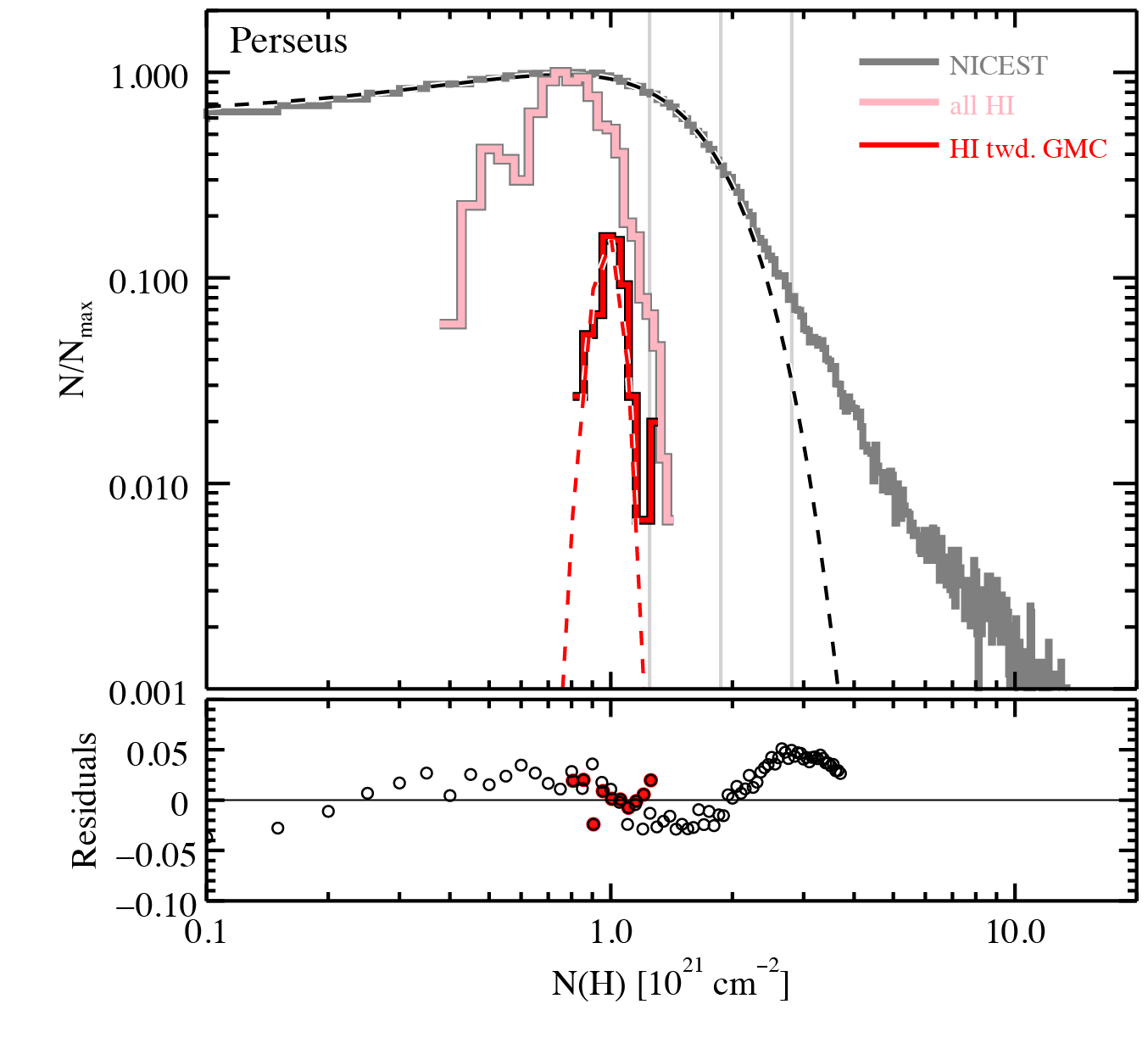} 
\caption{Same as Figure 8, but for the Perseus MC. \label{fig9}}
\end{figure}

\begin{figure}
\centering
\includegraphics[width=0.45\textwidth]{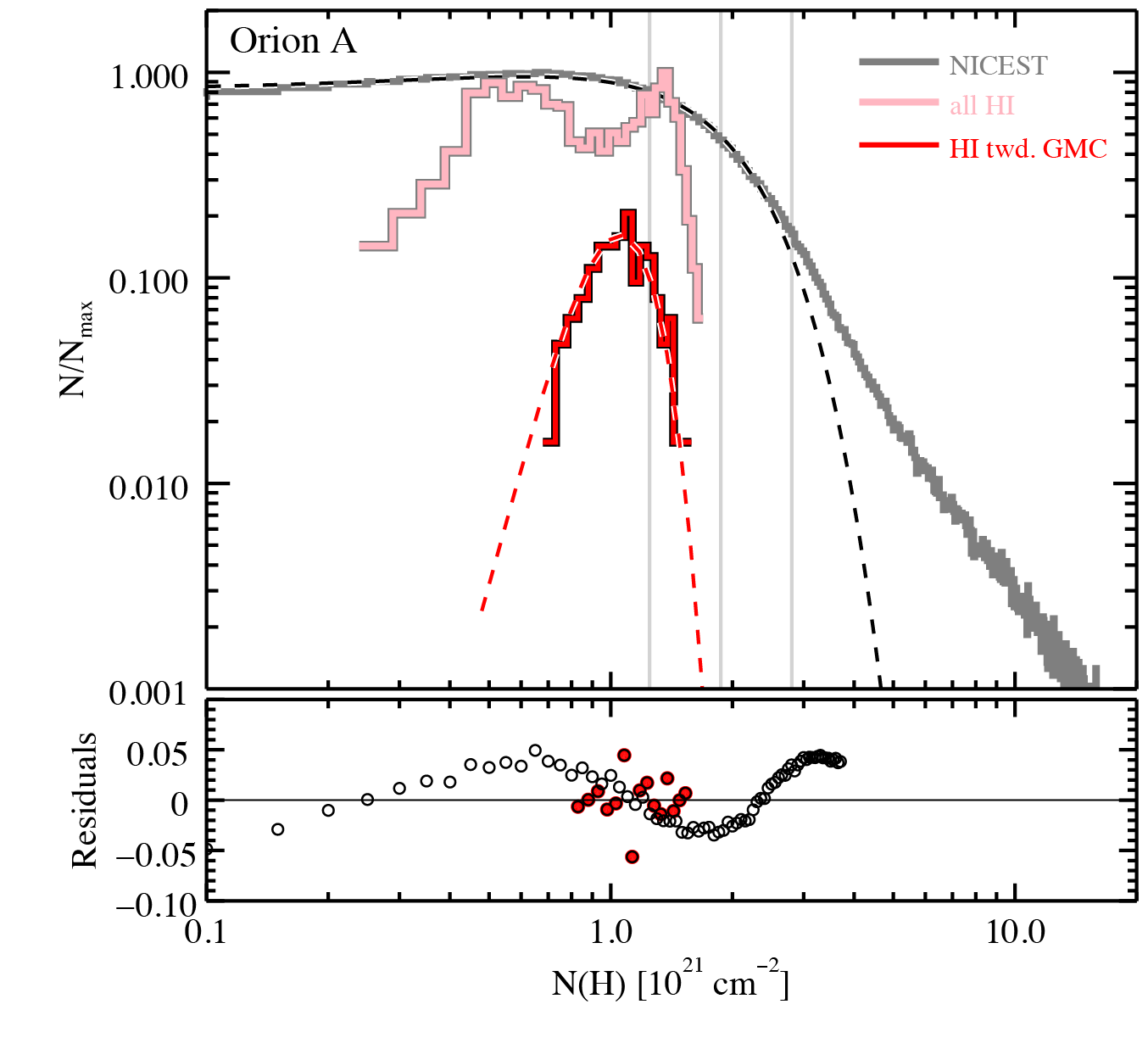}
\caption{Same as Figure 8, but for the Orion A MC.  \label{fig10}}
\end{figure}

\begin{figure}
\centering
\includegraphics[width=0.45\textwidth]{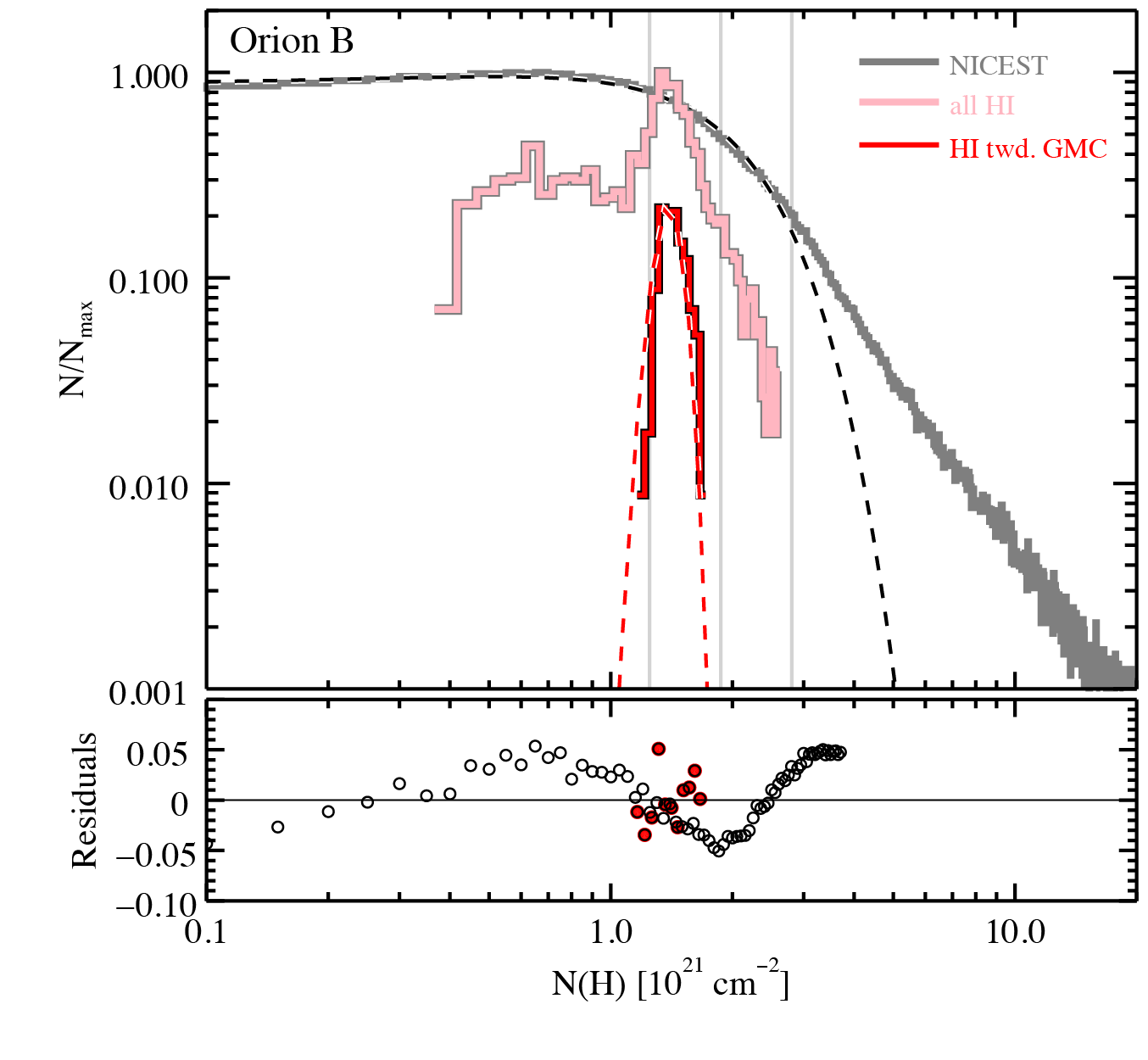}
\caption{Same as Figure 8, but for the Orion B MC.  \label{fig11}}
\end{figure}

\begin{figure}
\centering
\includegraphics[width=0.45\textwidth]{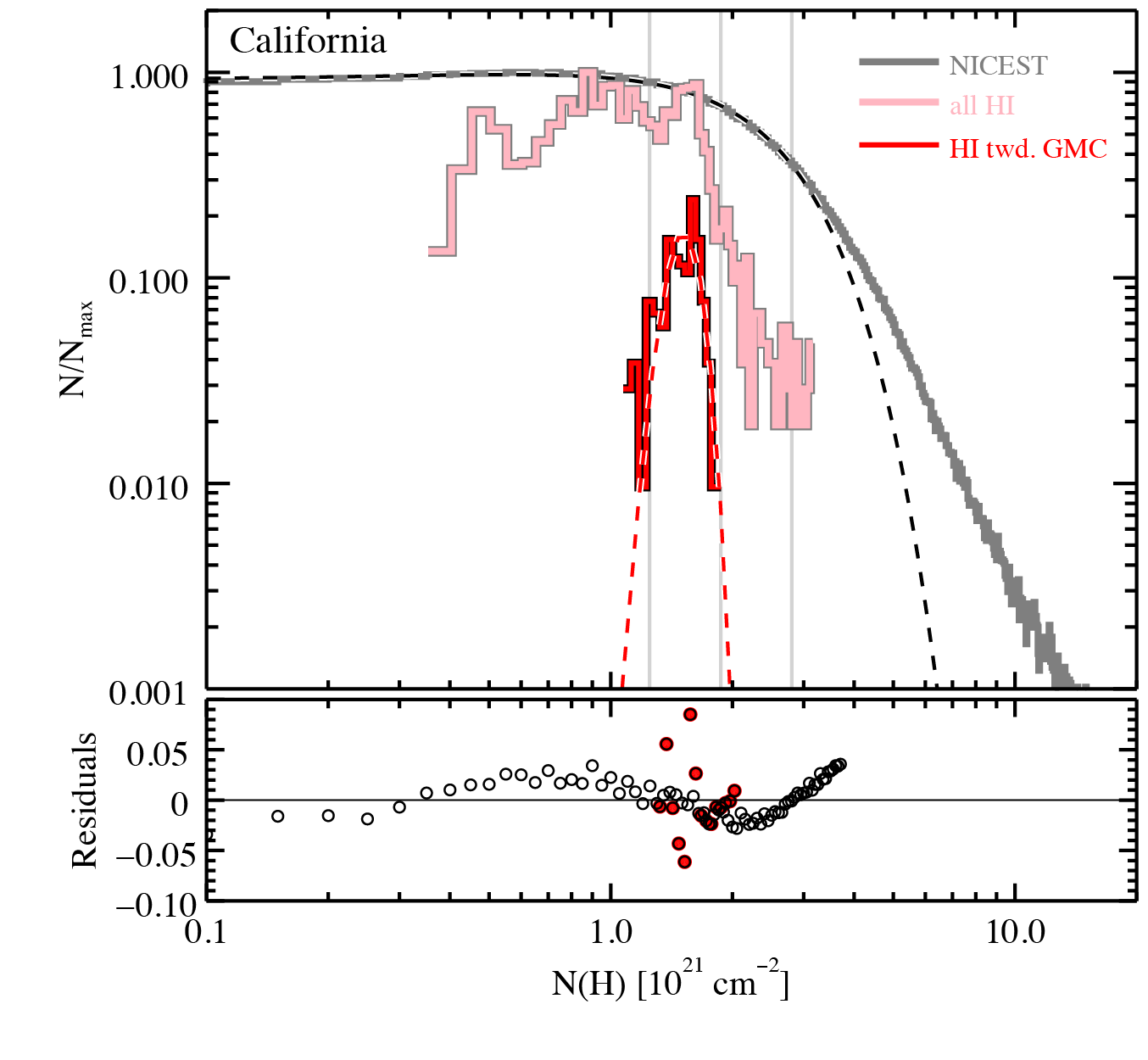} 
\caption{Same as Figure 8, but for the California MC.  \label{fig12}}
\end{figure}

\begin{figure}
\centering
\includegraphics[width=0.45\textwidth]{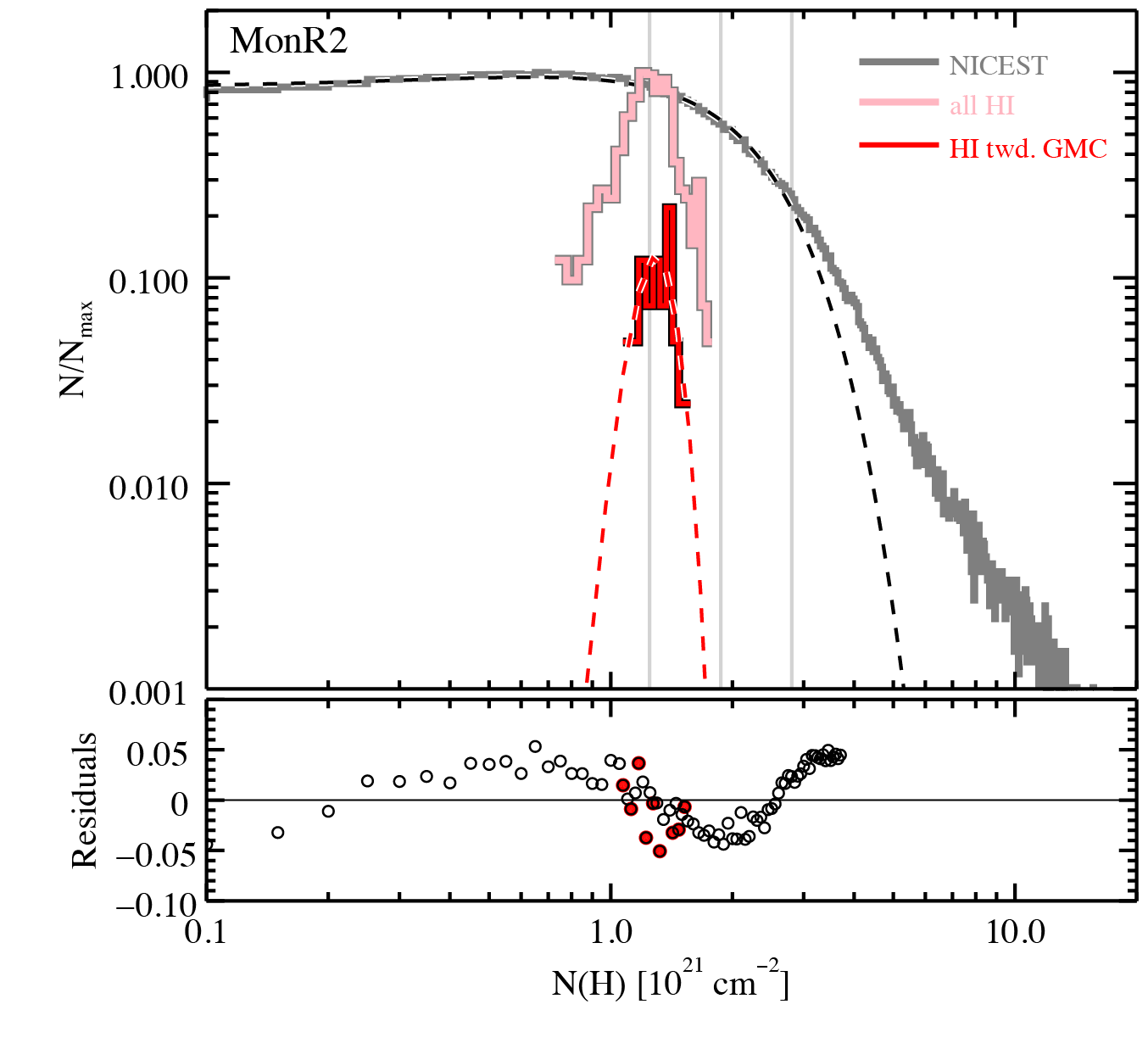}
\caption{Same as Figure 8, but for the MonR2 MC.  \label{fig13}}
\end{figure}

\begin{figure}
\centering
\includegraphics[width=0.45\textwidth]{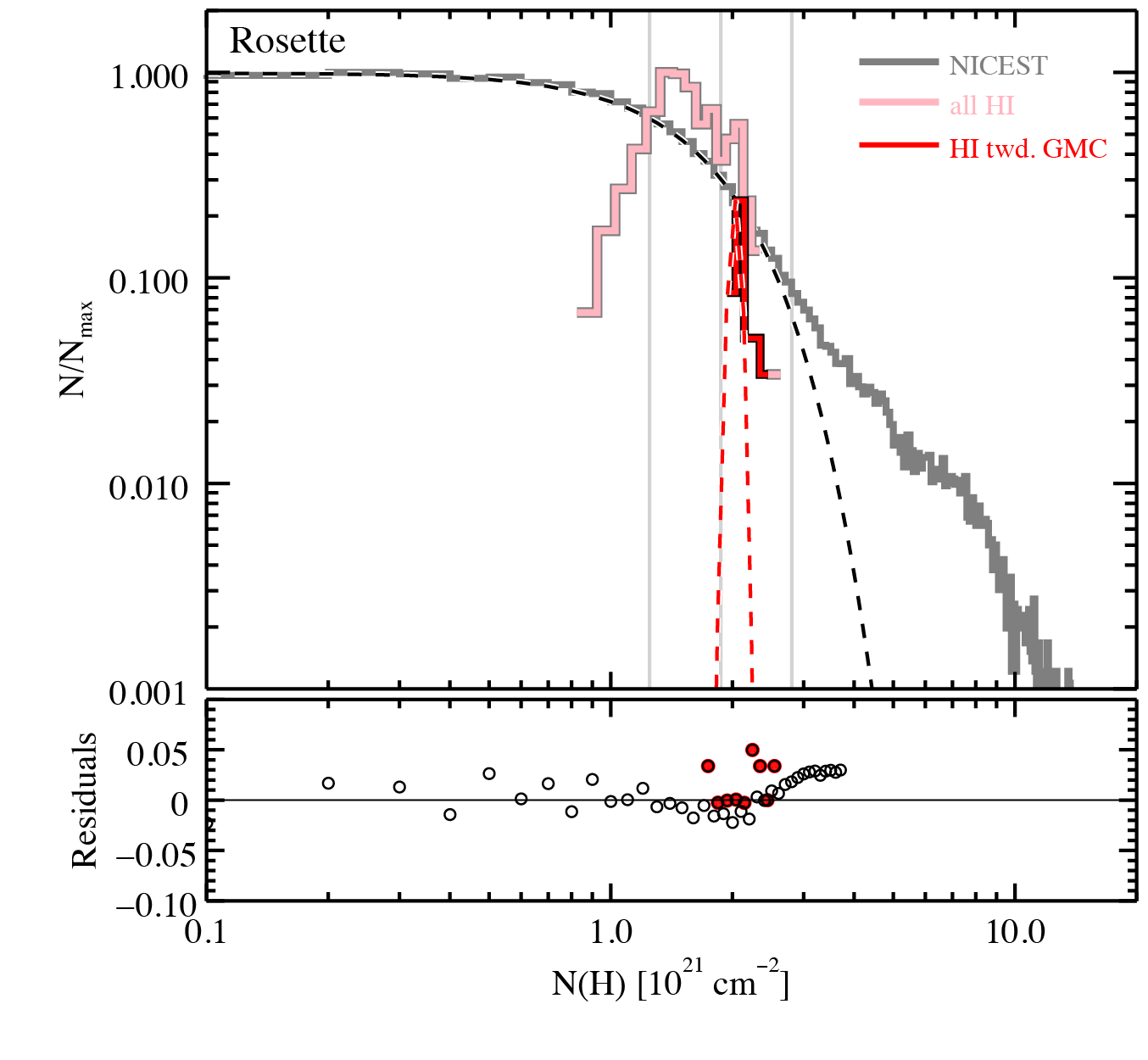}
\caption{Same as Figure 8, but for the Rosette MC, and the bin size is $0.1\times 10^{21}~\cm$.  \label{fig14}}
\end{figure}

The \HI~PDFs towards MCs shown in Figures \ref{fig8}--\ref{fig14} (red curves) have a number of noteworthy features.  To begin with, the peaks of the \HI~PDFs, summarized in Table \ref{table2}, fall in the range $\nhi \approx 1$ -- $2\times 10^{21}~\cm$.   Perseus, Orion A, Orion B, California, and MonR2---clouds laying within $\sim 500$ pc of one another and located well-below the Galactic plane where the problem of blending along the line of sight is mitigated---have a tighter range of peak column densities, within 1 -- $1.5\times 10^{21}~\cm$.   At a distance of about 1,330 pc and a Galactic latitude of $b_0=-2^\circ$, the Rosette is the most distant molecular cloud in our sample as well as the closest to the Galactic plane, where confusion due to unrelated components along the line of sight is exacerbated.  Thus, the value of $\nhi=(2.03\pm 0.03) \times 10^{21}~\cm$ for the Rosette may be an upper limit of the peak \HI~column density.  Indeed, the \HI~spectra in Figure \ref{fig0} show that, compared to the other clouds, the Rosette is associated with the most 21 cm emission by far.  Moreover, there is a possibly unrelated component to the emission at $\sim 20~\kms$ that is difficult to disentangle.
  
If we apply the fiducial gas-to-dust ratio to \nhi,  our results show that the peak of the distribution of \HI~column densities is typically somewhat less than the \HI-to-\htwo~transition of 1 mag predicted by Krumholz \& McKee (2009) for Solar metallicity clouds.  Yet the precise location of the peak has some variability.

\begin{figure}[t]
\plotone{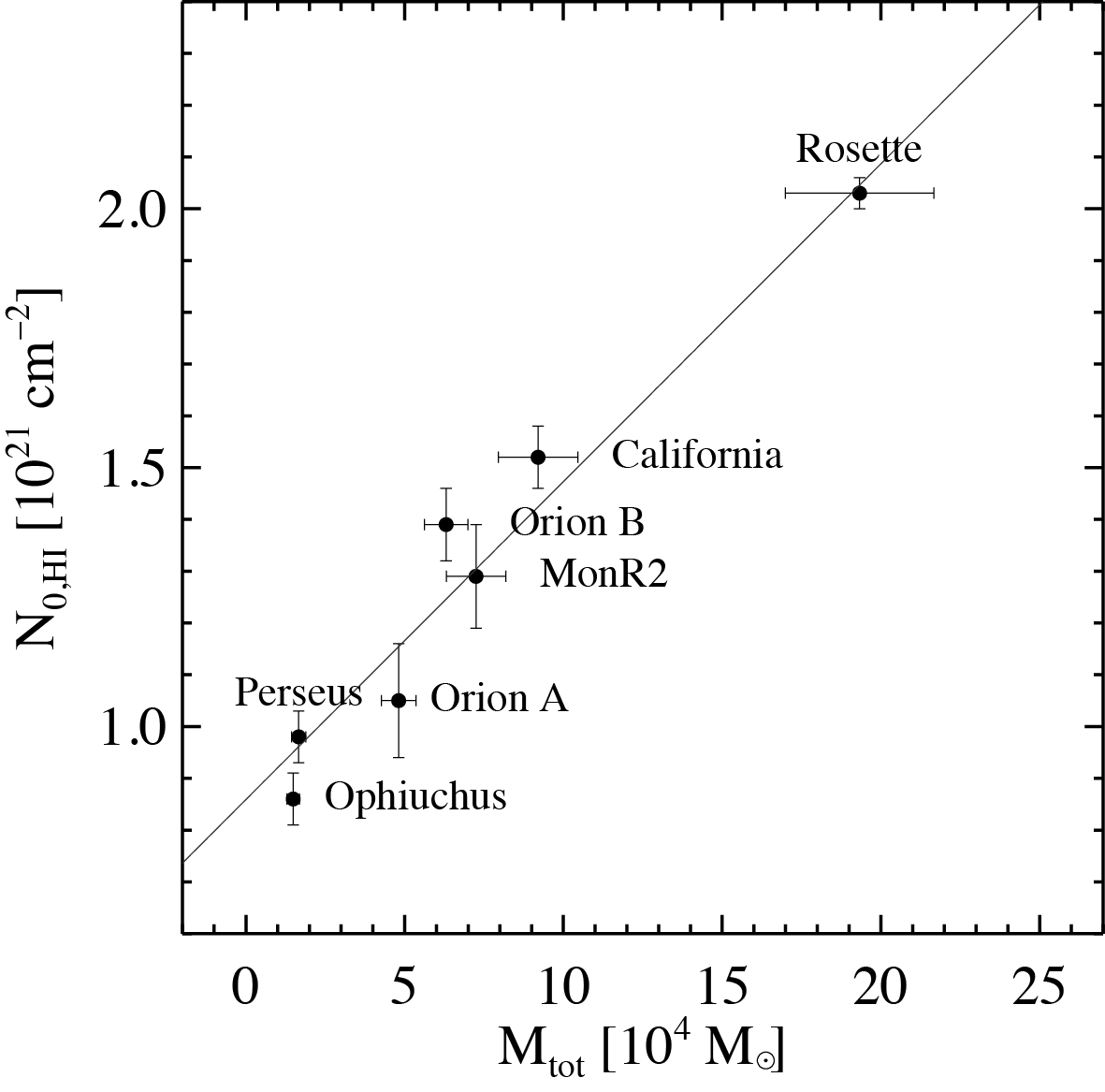} 
\caption{Peak column density of the \HI~PDF, $N_{0,\rm HI}$, versus total cloud mass, $M_{\rm tot}$, from the NICEST data. Overplotted is a least-squares fit to the data: $N_{0,\rm HI}/10^{21}~\cm = (0.061\pm 0.002) (M_{\rm tot}/10^4\msun) + (0.86\pm 0.03)$. \label{fig15}}
\end{figure}

The variations in $\nhi$ may be related to a correlation with total cloud mass, $M_{\rm tot}$.  As Figure \ref{fig15} demonstrates, $N_{0,\rm HI}$ increases with the total gas mass towards cloud complexes, $M_{\rm tot}$, as derived from the NICEST data.  In estimating the errors for $M_{\rm tot}$, we assume a 10\% 1-$\sigma$ uncertainty in the fiducial gas-to-dust ratio.   A least-squares fit to the data yields $N_{0,\rm HI}/10^{21}~\cm=(0.061\pm 0.002)(M_{\rm tot}/10^4\msun)+(0.86\pm 0.03)$.  Imara et al. (2011) found a similar trend for \HI~regions associated with MCs in M33; they found that the \HI~surface density scales as $M_{\rm CO}^{0.27}$, where $M_{\rm CO}$ is the molecular cloud mass as determined from CO observations.   That \nhi~increases with $M_{\rm tot}$ implies that high-mass MCs also tend to be associated with more \HI~overall.  We calculate the fraction of the mass in each cloud corresponding to \HI, according to $f_{\rm HI}=M_{\rm HI}/M_{\rm tot}$, where the \HI~mass towards the MC, $M_{\rm HI}$, is measured from the LAB data, and $M_{\rm tot}$ is the total mass of gas above $A_{\rm V,NICEST}\ge 1$ mag as measured from the NICEST extinction map.  As summarized in Table \ref{table2}, $f_{\rm HI}$ ranges from 0.25 (for Ophiuchus) to 0.51 (for the Rosette).  The four highest mass MCs in our sample (Orion B, California, MonR2, and the Rosette) have \HI~mass fractions $f_{\rm HI}\gtrsim 0.4$; the three lower mass clouds (Ophiuchus, Perseus, and Orion A) have lower \HI~mass fractions of $f_{\rm HI}< 0.3$.
 
We do not see a trend between  $f_{\rm HI}$ and the level of star formation activity, as measured from the population of young stellar objects (YSOs) in each molecular cloud.  Lada et al. (2010) assumed that the number of YSOs, $N_{\rm YSOs}$, is directly linked to the star formation rate (SFR) in a cloud to derive $\rm{SFR} [M_\odot \rm{Myr}^{-1}]=0.25N_{\rm YSOs}$.  Using this relation, the California MC, with 279 YSOs, has the lowest SFR of the MCs in our sample (see Table \ref{table2}).  Yet California has a relatively high \HI~mass fraction of $f_{\rm HI}=0.38$, similar to that of other high-mass clouds with many more YSOs.  Perseus and Ophiuchus have around 319 and 37 more YSOs than California, respectively, but they have lower \HI~mass fractions.  
 
As discussed in Chromey et al. (1989) and Elmegreen (1989), atomic clouds with high densities can be due to the presence of a high-intensity UV radiation field in the vicinity.  In a number of cases, the MC sits at the edge of a great deal of high-column-density \HI.  In particular, the PDFs of Ophiuchus, Orion A, Orion B, and California reveal that the \HI~outside the boundaries of the MCs, and within the accumulation radius of $\sim 3R_{\rm MC}$, reaches maximum column densities of $N(\HI)\gtrsim 1.5\times 10^{21}~\cm$, higher than the columns inside the boundaries of the MCs.   It is instructive to think about why we do not see molecular clouds in these high-column-density regions since, presumably, the column densities are high enough for \HI-dust shielding to occur, allowing for the efficient formation of \htwo.  To north of Orion A in the vicinity of the nebula, for instance, there is a high-density \HI~ridge, also recognized by Chromey et al. (1989).  They pointed out that this high-density feature, the $\sigma$ Ori shell (located near $l=207^\circ, b=-17\fdg 3$ in Figure \ref{fig3}), may be in transition between the \htwo~and H$^+$ phases.  This high pressure structure is also likely to have a strong UV radiation field, and has not had time to expand away from the MC into the \HII~region.  Since a correlation between \nhi~and dust temperature towards clouds is anticipated if variations in \nhi~depend on the UV radiation field, a future study could compare \nhi~with dust temperature observations, for instance, from the \emph{Planck} Collaboration (Planck Collaboration et al. 2014).  

In at least a few other cases, such as Ophiuchus (Figure \ref{fig1}), California (Figure \ref{fig5}), and MonR2 (Figure \ref{fig6}), there is high-column-density \HI~that is so far removed from the NICEST part of the MC, and with no \HII~region in the vicinity, it is difficult to see how this gas could be a transient feature associated with the MC itself.  Perhaps the combination of other physical conditions in these regions---such as dust content, magnetic fields, turbulence, and temperature---make them ill-suited for the formation of molecular clouds.  Alternatively, these high-density \HI~regions may be the precursors of future generations of molecular clouds.

\begin{figure}[t]
\plotone{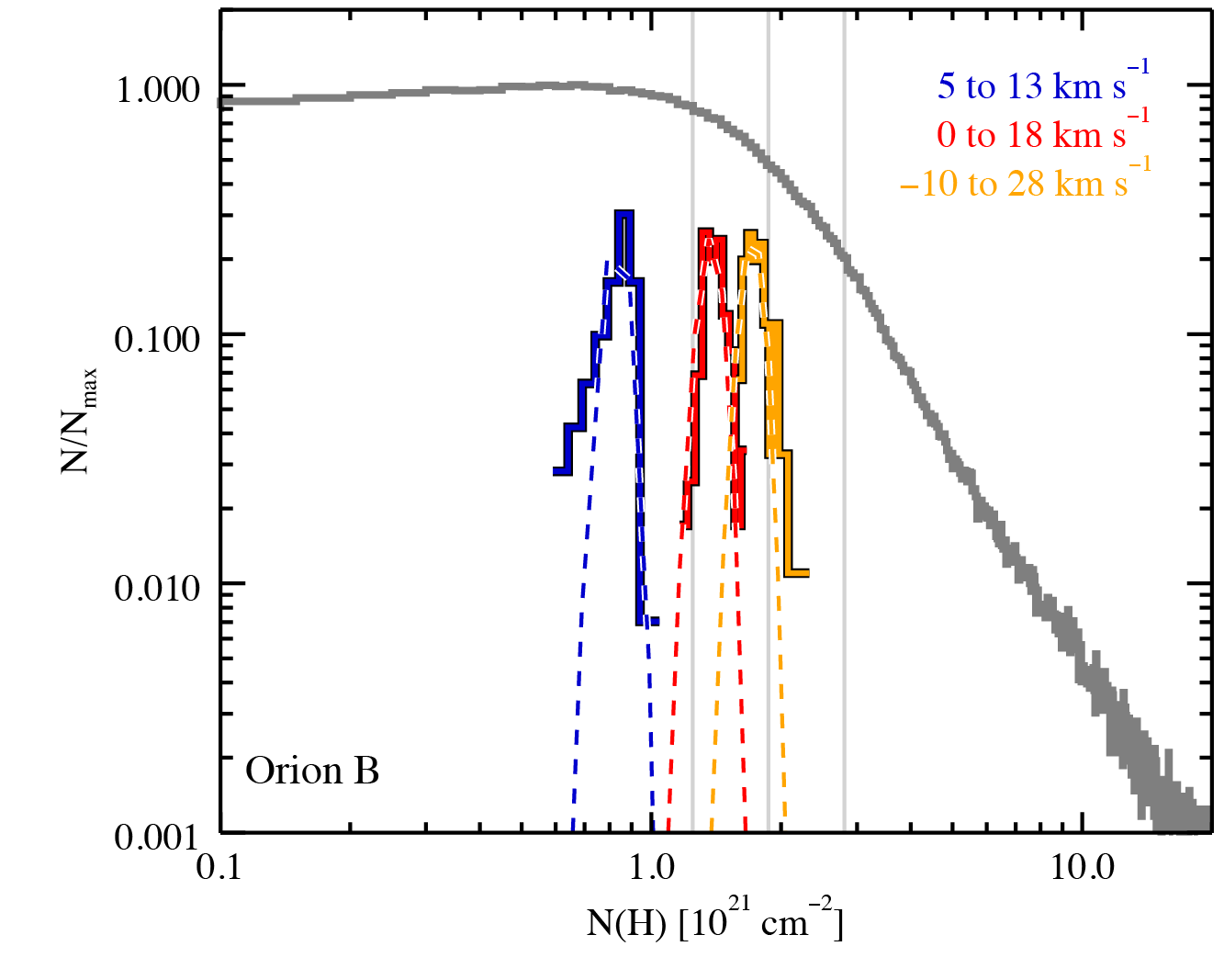} 
\caption{The Orion B \HI~PDFs for three velocity ranges.  The NICEST and \HI~PDFs are normalized as in Figure \ref{fig1}.  \label{fig16}}
\end{figure}

\subsection{Low-column-density \HI}\label{sec:lowdensity}
Toward the MCs in our sample, as indicated by the red curves in Figures \ref{fig8} -- \ref{fig14}, there is a dearth of low-column-density \HI~with $N(\HI)\lesssim 0.6-1\times 10^{21}~\cm$.   This stands in contrast to the NICEST PDFs (gray curves), which---taking into account all of the gaseous material, atomic and molecular, toward the clouds---seem to reveal an excess of low-column-density material.  By contrast, most of the low-density \HI~resides in the larger regions surrounding the MCs (pink curves in Figures \ref{fig8} -- \ref{fig14}).  Of the clouds in this sample, the \HI~PDFs toward Ophiuchus, Perseus, and Orion A (red curves) exhibit the most amount of low-column-density material, with $N(\HI) < 1\times 10^{21}~\cm$.  

The dispersions of the \HI~PDFs, $\sigma_{\rm HI}$, are uniformly narrow, averaging $0.12\times 10^{21}~\cm$.  Orion A, with $\sigma_{\rm HI}=(0.20\pm 0.12)\times 10^{21}~\cm$,  has the broadest PDF in the sample.  In their study of the atomic gas in the Orion region, Chromey et al. (1989) noted the asymmetrical, ``windswept'' appearance of the \HI~emission in the region, arising due to the marked density gradient from east to west (in Equatorial coordinates).  This asymmetry is also strikingly apparent in Galactic coordinates as a north-south gradient (Figure \ref{fig3}).  This dramatic density gradient may partially explain the relatively wide dispersion of the Orion A \HI~PDF.

In the following, we discuss how adjusting the kinematic and spatial boundaries we used to select \HI~emission (see \S \ref{sec:maps}) impacts the shape of the PDF.  The purpose of this exercise is to gauge the robustness of our results, especially given our limited knowledge about what \HI~emission ``belongs'' to a cloud complex.  First, we assess how integrating over different velocity ranges of \HI~emission modifies the PDF.  We compare three different velocity ranges: a ``normal'' range (those listed in Table \ref{table1}), a narrow range, and a wide range.  For the narrow range, we shift the minimum (maximum) velocities listed in Table \ref{table1} inward, by $+5~\kms$ ($-5~\kms$).  We also integrate over a wide range of velocities, by expanding the minimum (maximum) velocities in Table \ref{table1} by $-10~\kms$ ($+10~\kms$).   Figure \ref{fig16} shows three \HI~PDFs for the atomic gas toward with Orion B, generated by integrating over 5 to 13 \kms~(narrow; blue curve), $0$ to 18 \kms (normal; red curve, as in Figure \ref{fig11}); and $-10$ to 28 \kms~(wide; orange curve).  The main consequence of decreasing the range of velocities is to shift the peak of the PDF to smaller values of $\nhi$.  But when changing the range of velocities over which one integrates, \emph{the narrow, log-normal shape of the \HI~PDF is retained}.  This trend holds for the other clouds in our sample, the summary for which is in Table \ref{table3}.  With few exceptions, the dispersions of the PDFs are identical, to within the 1-$\sigma$ margin of error, as the velocity range around the \HI~line center is changed by a few to tens of \kms.

What happens to the \HI~PDF when one varies the spatial boundaries of the selected \HI~but keeps the velocity range fixed?  In the NICEST maps, relaxing the extinction threshold to a smaller value of \av~amounts to defining clouds with larger spatial areas.  We quantify how the peak of the \HI~PDF change for three different boundaries: within the $A_{\rm V,NICEST}\ge 1.5$, $\ge 1$, and $\ge 0.5$ mag levels of the NICEST extinction maps.  Table \ref{table4} summarizes the results and lists the number of pixels contained within each of the boundaries.  The table shows that when the spatial boundary defining an MC is increased ($A_{\rm V,NICEST}$ decreases), the general trend is for the peak of the PDF to shift by small amounts toward lower column densities.  Expanding the spatial boundaries by $\sim 0.5$ mag to include more atomic gas does not result in an excess of low-column-density material below the peak of the \HI~PDF.  Moreover, as with varying the velocity range, the log-normal shape of the PDF is generally preserved.  In most cases, $\nhi$ and $\sigma_{\rm HI}$ remain the same, to within the 1-$\sigma$ margin of error, leading us to conclude that our choice of the 1 mag contour is appropriate and that our results are robust. 
  
A comparison of the extinction map PDFs and \HI~PDFs towards MCs, (gray and red curves, respectively, in Figures \ref{fig8} -- \ref{fig14}), shows that most of the low-column-density material towards MCs cannot be explained by warm, diffuse atomic hydrogen.  The origin of the low-extinction material towards molecular clouds has been a concern of previous studies that measured PDFs from dust extinction or dust emission observations (e.g., Kainulainen et al. 2009; Lombardi et al. 2011; Imara 2015).  Particularly, it is challenging to disentangle low-extinction\footnote{We will speak of column density, extinction, and surface density interchangeably, depending on the context.  These quantities are related by the conversions provided in equations (\ref{eq:gdr}) and (\ref{eq:surface}).} material arising from a given cloud itself from other effects, including foreground and background contamination along the line of sight to the cloud, projection effects, and unsuitable delineation of the cloud boundaries.  For these reasons, Lombardi et al. (2015) argued that low-extinction measurements, below $\av\lesssim 1$ mag, from dust extinction or emission maps are unreliable. 

\subsection{The \HI-to-\htwo~Transition}\label{sec:transition}
The transitioning of atomic to molecular hydrogen is a critical first step in the chain of events leading to star formation.  Once a cold atomic cloud becomes sufficiently well shielded against dissociating UV photons from the interstellar radiation field, molecular hydrogen can form, and star formation is possible.  Given the many functions of \HI~in MC and stellar evolution highlighted in the introduction, the \HI-to-\htwo~transition has been the subject of a number of observational studies. In the Galaxy, some authors determined \htwo~column densities from UV absorption measurements and compared these with \HI~21-cm observations (Savage et al. 1977; Gillmon et al. 2006).  Other studies used indirect measurements of \htwo---far-infrared, dust extinction, or CO observations, for instance---to infer the \HI-to-\htwo~transition in the Milky Way (e.g., Reach et al. 1994; Meyerdierks \& Heithausen 1996; Lee et al. 2012) and external galaxies (e.g., Wong \& Blitz 2002; Blitz \& Rosolowsky 2004, 2006; Bigiel et al. 2008).

Krumholz et al. (2008) investigated the \HI-to-\htwo~transition by modeling a spherical cloud bathed in a uniform, isotropic radiation field.  They later applied their model to galactic observations and showed that the \HI-to-\htwo~transition occurs at a characteristic \HI~column density of $\sighi\approx 10~\sunits$ in Solar metallicity clouds (Krumholz et al. 2009; hereafter, KMT09).  This column is in close agreement with observations of $\sighi$ towards extragalactic molecular clouds (e.g., Wong \& Blitz 2002; Blitz \& Rosolowsky 2004; Imara et al. 2011) and Galactic molecular clouds (e.g., Lee et al. 2012). In this section, we show that the surface density (i.e., column density) at which the peak of the \HI~PDF occurs is in close agreement with the \HI~surface density at the \HI-to-\htwo~transition predicted by KMT09.

To compare our results with KMT09 and other studies, we first convert \HI~and \htwo~column densities to units of surface density, \sighi~and \sightwo,  using
\begin{equation}\label{eq:surface}
\frac{\sighi}{\sunits} =\frac{N(\HI)}{1.25\times 10^{20}~\cm}.
\end{equation}
We follow KMT09 and Lee et al. (2012) in deriving an \htwo-\HI~ratio, defined as 
\begin{equation}\label{eq:rh2}
\rhtwo \equiv\frac{\sightwo}{\sighi} = \frac{\Sigma_{\rm H}}{\sighi}-1,
\end{equation}
where  $\Sigma_{\rm H}=\sightwo + \sighi$ is the total hydrogen surface density at each location in a cloud complex.  

In the following, we consider the \HI~surface density within the accumulation radius of each MC, (as defined in \S \ref{sec:maps}), \emph{not} just within the $\av=1$ mag contour.  This is because we want to investigate the surface density at which the \HI-to-\htwo~transition occurs without any \emph{a priori} assumptions.  In the vicinity of Ophiuchus, \sighi~ranges from roughly 2 -- 18 \sunits; in Perseus, $\sighi\approx 3-11$ \sunits; in Orion A, $\sighi\approx 2-13$ \sunits; in Orion B, $\sighi\approx 3-16$ \sunits; in California, $\sighi\approx 3-26$ \sunits; in MonR2, $\sighi\approx 6-14$ \sunits; and in the Rosette, $\sighi\approx 7-21$ \sunits.

Our results for Perseus are in  good agreement with Lee et al. (2012), who found that \HI~surface density of $\sighi\approx 5-11$ \sunits~across the cloud.  These authors examined the \HI-to-\htwo~transition in Perseus using $\sim 9$ times \emph{higher resolution} 21-cm observations ($\sim 4^\prime$) from the Galactic Arecibo \emph{L}-band Feed Array \HI~Survey (GALFA-\HI~Survey; Peek et al. 2011).  Most likely, this difference in angular resolution does not give rise to discrepancies in the derived \HI~surface densities of our respective studies because, as Stanimirovi\'{c} et al. (2014) demonstrated, the diffuse, WNM dominates over the clumpier CNM across Perseus.

For a spherical, Solar metallicity cloud embedded in a uniform radiation field, KMT09 predicted $\sighi\approx 10~\sunits$ at the transition between the atomic- and molecular-dominated regions of the cloud.  They determined that \rhtwo~is approximately 25\% at the transition.  For each cloud in our sample, we calculate \sighi~at $\rhtwo=0.25\pm 0.1$ and report the results in Table \ref{table5}. The table also lists values of $\Sigma_{0,\rm HI}$, the location of the peak of the \HI~PDFs in units of surface density (defined at $\av\ge 1$ mag).  Bearing in mind that \rhtwo~depends on the chosen gas-to-dust ratio through \sightwo, via equations (\ref{eq:gdr}) and (\ref{eq:surface}), we find that $\sighi(\rhtwo=0.25)$ is remarkably consistent with $\Sigma_{0, \rm HI}$.  Ophiuchus, for instance, has an \HI~surface density of $7.4\pm 1.3$ \sunits~at $\rhtwo=0.25$, and from the fit to its \HI~PDF, $\Sigma_{0, \rm HI}=6.9\pm 0.5~\sunits$.  

The transition column density we measure for Perseus, $6.8\pm 1.4$, is again in excellent agreement with Lee et al. (2012), who estimated \HI~surface densities of $\sim 6$--8 \sunits~towards both dark and star-forming regions in the cloud.  But to derive $N(\htwo)$ and, therefore, \sightwo~from their extinction maps of Perseus, Lee et al. (2012) applied a gas-to-dust ratio of $9.1\times 10^{20}$ mag$^{-1}~\cm$, $\sim 1/2$ the typical Galactic ratio we adopt here (see equation \ref{eq:gdr}).  These authors obtained this ratio by comparing \av~with their high-resolution measurements of $N(\HI)$ around Perseus.  If we instead use the Lee et al. gas-to-dust ratio to determine \rhtwo, we measure $\sighi(\rhtwo=0.25)=6.2\pm 1.7$, still in excellent agreement with their conclusions.  This is because applying a different gas-to-dust ratio to derive \sightwo~is equivalent to calculating \sighi~at a slightly different value of \rhtwo.  And over a range of \rhtwo, \sighi~is nearly constant, since once \HI~begins to fully convert into \htwo, \sighi~saturates, even while \sightwo~increases.

For the remaining clouds---Orion A, Orion B, California, MonR2, and the Rosette---$\sighi(\rhtwo=0.25)$ is also consistent with $\Sigma_{0, \rm HI}$.  All in all, these results suggest the \HI~PDF may be used as a simple tool to quickly identify the location of the \HI-to-\htwo~transition in molecular clouds.

\begin{figure}[t]
\plotone{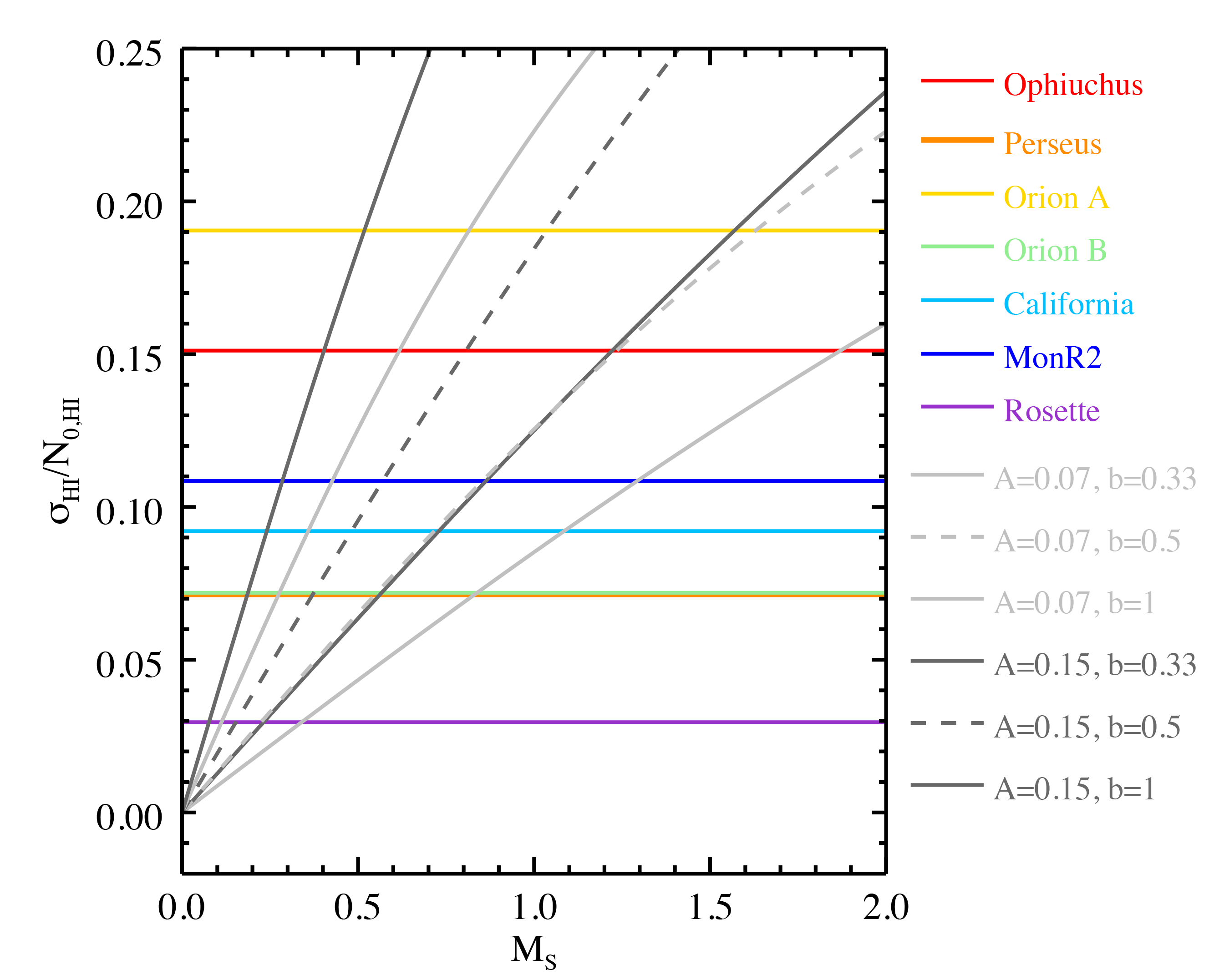} 
\caption{$\sigma_{\rm HI}/\nhi$ as a function of the sonic Mach number, $M_S$, for different values of $A$ and $b$ (see equation \ref{eq:mach}).  The solid horizontal lines indicate the location of $\sigma_{\rm HI}/\nhi$ for each cloud.  \label{fig:mach}}
\end{figure}

While the observed \sighi~saturation across the molecular clouds likely results from the \HI-to-\htwo~transition, it could alternatively be due to a large amount of the cold \HI~and \HI~self-absorption. The issue of CNM fraction and \HI~optical depth for the Perseus MC has been addressed in previous works (Lee et a. 2015 and Stanimirovi\'c et al. 2014). To conduct similar studies for each of the clouds here using absorption line measurements is well beyond the scope of this current paper.  However, the initial studies on the Perseus cloud provide a basis of comparison with the clouds in our sample which share similar \HI~properties to Perseus, such as similar \HI~PDF widths and cloud parameters. In the case of Perseus, Stanimirovi\'c et al. (2014) found that only 15\% of the lines of sight which had absorption line data had spin temperatures lower than 50 K.  They also found a rather low CNM fraction, between 10\%--50\%, implying that the WNM seems to be well mixed with the CNM throughout Perseus.  Lee et al. (2015) estimated a correction for the effects of self-absorption of \HI~in their derived Perseus map and found that optically thick \HI~only accounts for $\sim 20\%$ of CO-dark gas, and Burkhart et al. (2015) found that the optically thick \HI~correction does not significantly affect the shape of the \HI~PDF.  This can also be inferred from a comparison of our \HI~PDF of Perseus using LAB data (with no \HI~self-absorption correction) and the \HI~PDF from GALFA data with \HI~self-absorption corrected in Burkhart et al. (2015), both of which have log-normal PDF shapes with similarly narrow log-normal widths, $\sigma_{\rm HI}$.  Using their assumed gas-to-dust ratio for Perseus, Burkhart et al. (2015) found a dispersion $(0.13\pm 0.08)\times 10^{21}~\cm$, consistent with our finding of $\sigma_{\rm HI}=(0.07\pm 0.05)\times 10^{21}~\cm$ for Perseus.

Additional absorption line measurements are required to determine if all the clouds in our sample are similar to Perseus in that the \HI-to-\htwo~transition dominates the \HI~saturation rather than optically thick or cold \HI.  The 21-SPONGE survey (Murray et al. 2014, 2015) will provide additional absorption line data using the VLA and will be able to further test the importance of \HI~self-absorption and constrain the fraction of cold \HI~in and around molecular clouds.

\subsection{Sonic Mach Numbers}
Log-normal column density PDFs have been observed in both the atomic (diffuse neutral and ionized) and molecular (star-forming) ISM (e.g., Vazquez-Semadeni 1994; Hill et al. 2008; Brunt 2010; Kainulainen et al. 2011; Burkhart et al. 2010; Burkhart \& Lazarian 2012; Molina et al. 2012; Kainulainen \& Tan 2013).  Studies of log-normal  PDFs of interstellar clouds from both observational and theoretical perspectives have noted a relationship between the sonic Mach number measured in the cloud and the width (variance or standard deviation) of the gas PDF (e.g., Padoan et al. 1997; Price et al. 2011; Federrath et al. 2008; Kowal, Lazarian, \& Beresnyak 2007; Burkhart \& Lazarian 2012 and references therein). If a cloud becomes self-gravitating, then the shape of the turbulence-induced log-normal PDF begins to form a power-law tail which is skewed toward the high density material  (Klessen 2000; Collins et al. 2012). Furthermore, the PDF was shown to be important for analytic models of star formation rates and initial mass functions (Padoan \& Nordlund 2002; Krumholz \& McKee 2005; Hennebelle \& Chabrier 2011; Padoan \& Nordlund 2011; Federrath \& Klessen 2012, 2013). 

The relationship between the observable column density PDF standard deviation and sonic Mach number (Burkhart \& Lazarian 2012, their Equation 4) retains the same form as that of the three-dimensional density field but with a scaling constant $A$:
\begin{equation}\label{eq:mach}
\left(\frac{\sigma_{\rm HI}}{N_{0,\rm HI}}\right)^2=(b^2{\cal M}_s^2+1)^A-1.
\end{equation}
Equation (\ref{eq:mach}) depends on the sonic Mach number as well as the parameters $A$, (a function of the optical depth; see Burkhart et al. 2013a), and $b$ (which depends on the type of turbulence driving; see Federrath et al. 2008). 

Given a fitted value of the standard deviation of the \HI~reported in Table \ref{table2}, we can compute the Mach number expected from such a relationship for our \HI~PDF sample. In Figure \ref{fig:mach} we plot curves of the expected standard deviation-sonic Mach number relationship given in equation (\ref{eq:mach}) for different values of $A$ and $b$. We overplot the measured values of ${}^{\sigma_{\rm HI}}/_{N_{0,\rm HI}}$ (horizontal lines) for the \HI~cloud sample. Although $A$ and $b$ are unknown for these clouds, the realistic parameter space reported in the literature ($A\approx 0.1$, $b\approx {}^1/_3-1$; i.e., Federrath \& Klessen 2012; Burkhart \& Lazarian 2012) would predict that the \HI~in MCs is generally subsonic to transonic. The predicted Mach numbers for a cloud, for different combinations of $A$ and $b$, are given by where the gray curves in Figure \ref{fig:mach} cross the measured value of  ${}^{\sigma_{\rm HI}}/_{N_{0,\rm HI}}$ for that cloud.  Thus for instance, for $A=0.07$ and $b=0.33$, (the light gray dotted curve), the Mach numbers predicted from equation (\ref{eq:mach}) are ${\cal M}_s=1.9$, 0.8, 2.5, 0.6, 1.1, 1.3, and 0.3 for Ophiuchus, Perseus, Orion A, Orion B, California, MonR2, and the Rosette, respectively.

Observational studies that have investigated the sonic Mach number of \HI~CNM gas in the diffuse medium, based on measurements of velocity linewidths and the spin temperature (Heiles \& Troland 2003; Burkhart et al. 2010; Burkhart et al. 2015; Chepurnov et al. 2015), suggest that the CNM is supersonic with sonic Mach numbers as high as 10.  Supersonic to transonic motion of the \HI~gas is also predicted based on shallow slopes of the density power spectrum and has been observed in \HI~clouds (Stanimirovi\'c et al. 1999; Stanimirovi\'c \& Lazarian 2001; Burkhart et al. 2010; Chepurnov et al. 2015). These observational studies have found that the larger the fraction of WNM, the lower the sonic Mach number in the \HI~gas (Chepurnov et al. 2015).  Numerical simulations, which include thermal instability, similarly find that the CNM is high supersonic while the WNM is transonic (e.g., Gazol \& Kim 2013).

In this study, however, we find that the \HI~PDF widths in our sample of MCs are very narrow in and around the cloud complexes. Similarly, Burkhart et al. (2015) found that the width of the \HI~PDF towards Perseus is too narrow to be described by supersonic turbulence and is consistent with subsonic or transonic turbulence. In the case of Perseus, spin temperature measurements (probing the CNM) are available via 21-cm absorption lines from Stanimirovi\'{c} et al. (2014). Burkhart et al. (2015) found that the PDF of Perseus is very narrow and that equation (\ref{eq:mach}) predicts a subsonic Mach number much lower than direct measurements from the spin temperature (a median value of $M_s = 4.0$).  These findings suggest that there may be substantial WNM components in and around MCs which would lower the average sonic Mach number in the cloud from the measured values from the CNM absorption line studies.  Indeed, as mentioned above in \S\ref{sec:transition}, Stanimirovi\'{c} et al. (2014) estimated that in Perseus the CNM makes up only 10\%--50\% of the \HI~filling fraction.  An alternative explanation is that the \HI~PDF width is truncated in MCs past the \HI-to-\htwo~transition due to the depletion of \HI. 

\section{Summary and Conclusion}
We have presented a detailed investigation of the PDFs of atomic gas associated with seven Galactic molecular clouds having a range of SFRs.  Since most of the MCs in our sample lay far enough from the Galactic plane so that blending along the line of sight towards the clouds is mitigated, it was possible to isolate their \HI~envelopes using 21-cm observations and make reliable measurements of the properties. We created column density maps of the \HI~associated with MCs, we analyzed the features of the \HI~PDFs, and we measured the column density of atomic gas at the \HI-to-\htwo~transition.  Our most salient results are summarized as follows.
\begin{enumerate}
\item The \HI~column density PDFs associated with MCs tend to have narrow, log-normal shapes.  The peaks of the PDFs occur at column densities ranging from $\sim 1$--$2\times 10^{21}~\cm$.  

\item The properties of the \HI~PDFs are fairly robust to moderate variations to the spatial criteria used to select the 21-cm emission associated with the cloud complexes.  The location of the column density at which an \HI~PDF peaks, \nhi, is sensitive to the chosen velocity range of the 21-cm emission used to create the PDF; however, the dispersion of the \HI~PDF, $\sigma_{\rm HI}$ is fairly insensitive to variations in the kinematic criteria.

\item The column density at the peak of the \HI~PDF, \nhi, tends to increase with increasing MC mass.  Moreover, higher mass MCs tend to have higher \HI~masses, $M_{\rm HI}$, and slightly higher \HI~mass fractions, $f_{\rm HI}$.  

\item Most of the low-extinction material below $\av\lesssim 1$ mag represented in the NICEST column density maps towards a given MC cannot be explained by warm, diffuse \HI~towards the cloud.  If this low-extinction material is not dominated by warm atomic gas or dark molecular gas, it is probably due chiefly to foreground or background that is unrelated to the cloud.

\item The typical dispersion of the \HI~PDFs is $\sigma_{\rm HI}\approx 0.12\times 10^{21}$ \cm. The sonic Mach numbers we would predict for the \HI~based on measurements of the PDF variance in clouds would be subsonic to transonic.  This is in contrast to the typical supersonic CNM sonic Mach number measurements found in both observations and simulations (e.g.,  Heiles \& Troland 2003;  Gazol \& Kim 2013; Burkhart et al 2015).  This suggests that there is a substantial WNM component in and around MCs which would lower the average sonic Mach number in the cloud, and/or the \HI~PDF width is truncated in MCs past the \HI-to-\htwo~transition due to the depletion of \HI.

\item At the \HI-to-\htwo~transition in cloud complexes, we estimate \HI~surface densities ranging from $\sim 7$--16 \sunits.  These values are consistent with $\Sigma_{0, \rm HI}$ obtained from least-squares fitting to the \HI~PDFs.  We propose that---assuming an appropriate selection criteria for selecting \HI~emission is adopted---the \HI~PDF is a useful tool for identifying the \HI-to-\htwo~transition column in Galactic MCs.
\end{enumerate}

\acknowledgements
We thank Tom Dame, Bruce Draine, and Charlie Lada for their helpful comments on earlier drafts of this paper; and we thank  Marco Lombardi for providing the NICEST data used in this study.  We thank the anonymous referee whose insights and thorough attention to an earlier draft helped to improve this paper.  N. Imara is supported by the Harvard-MIT FFL Fellowship, and B.B. is supported by the NASA Einstein Postdoctoral Fellowship.

\clearpage

\begin{table}\centering
\begin{center}
\begin{tabular}{lccccc}
\multicolumn{6}{c}{Table 2: \HI~PDF Properties \& \HI~Masses.}\\
\tableline\tableline
Cloud      & $N_{\rm 0,HI}$ [$10^{21}~\cm$] &  $\sigma_{\rm HI}$ [$10^{21}~\cm$]  & $M_{\rm HI}~[\msun]$ & $f_{\rm HI}$ & $N_{\rm YSOs}$   \\  
\tableline
Ophiuchus (1)  & $0.86\pm 0.05$   &  $0.13\pm 0.06$   & $3722$  &  $0.25\pm 0.04$    & 316    \\
Perseus (2)    & $0.98\pm 0.05$   &  $0.07\pm 0.05$   & $7983$  &  $0.27\pm 0.04$    & 598    \\
Orion A (3)    & $1.05\pm 0.11$   &  $0.20\pm 0.12$   & $12,511$ &  $0.26\pm 0.04$   & 2862  \\  
Orion B (4)    & $1.39\pm 0.07$   &  $0.10\pm 0.07$   & $25,225$ &  $0.40\pm 0.05$   & 635    \\  
California (5) & $1.52\pm 0.06$   &  $0.14\pm 0.07$   & $34,973$ &  $0.38\pm 0.06$   & 279    \\
MonR2 (6)      & $1.29 \pm 0.10$  &  $0.14\pm 0.11$   & $24,996$ &  $0.41\pm 0.06$   & 1003  \\  
Rosette (7)    & $2.03 \pm 0.03$  &  $0.06\pm 0.02$   & $98,589$ &  $0.51\pm 0.06$   & 461    \\ 
\tableline
\end{tabular}
\caption{The extinction at the peak of the \HI~PDF, $\nhi$, obtained from a Gaussian fit to the PDF; the dispersion $\sigma_{\rm HI}$ of the Gaussian fit to the PDF; the \HI~mass, $M_{\rm HI}$, associated with the MC and the \HI~mass fraction, $f_{\rm HI}$.  The $1\sigma$ uncertainties to the fits are quoted for $N_{\rm 0,HI}$ and $\sigma_{\rm HI}$.  
\textbf{YSO References.} (1) Wilking et al. (2008); (2) Lada et al. (2006); (3) Allen \& Davis (2008); (4) Lada et al. (1991); (5) Wolk et al. (2010); (6) Gutermuth et al. (2011); (7) Ybarra et al. (2013). }\label{table2}
\end{center}
\end{table}

\begin{table}\centering
\begin{center}
\begin{tabular}{lccc}
\multicolumn{4}{c}{Table 3: $N_{\rm 0,HI}$ for different velocity ranges.}\\
\tableline\tableline
Cloud       & $N_{\rm 0,HI}$ [$10^{21}~\cm$]      &  $N_{\rm 0,HI}$ [$10^{21}~\cm$]        &  $N_{\rm 0,HI}$ [$10^{21}~\cm$]  \\
            & $\sigma_{\rm HI}$ [$10^{21}~\cm$]   &  $\sigma_{\rm HI}$ [$10^{21}~\cm$]     &  $\sigma_{\rm HI}$ [$10^{21}~\cm$]  \\

            & \footnotesize{for $\pm 5~\kms$} &  \footnotesize{for $\pm 0~\kms$}  &  \footnotesize{for $\pm~10\kms$} \\  
\tableline
Ophiuchus   &  $0.25\pm 0.03$    &  $0.86 \pm 0.05$   & $1.22 \pm 0.09$  \\
            &  $0.04 \pm 0.03$   &  $0.13 \pm 0.06$   & $0.22 \pm 0.10$  \\
Perseus     &  $0.74 \pm 0.05$   &  $0.98 \pm 0.05$   & $1.06 \pm 0.05$  \\
            &  $0.06 \pm 0.05$   &  $0.07 \pm 0.05$   & $0.08 \pm 0.05$  \\
Orion A     &  $0.57 \pm 0.10$   &  $1.05\pm 0.11$    & $1.39 \pm 0.08$  \\  
            &  $0.18 \pm 0.12$   &  $0.20\pm 0.12$    & $0.23 \pm 0.08$  \\
Orion B     &  $0.84 \pm 0.01$   &  $1.39\pm 0.07$    & $1.75 \pm 0.03$  \\  
            &  $0.04 \pm 0.02$   &  $0.10\pm 0.07$    & $0.12 \pm 0.03$  \\
California  &  $0.86 \pm 0.07$   &  $1.52\pm 0.06$    & $2.10 \pm 0.07$  \\
            &  $0.14 \pm 0.08$   &  $0.14\pm 0.07$    & $0.23 \pm 0.07$  \\
MonR2       &  $0.56 \pm 0.03$   &  $1.29 \pm 0.10$   & $1.89 \pm 0.10$  \\
            &  $0.05 \pm 0.03$   &  $0.14\pm 0.11$    & $0.18 \pm 0.11$  \\  
Rosette     &  $0.62 \pm 0.05$   &  $2.03 \pm 0.03$   & $3.43 \pm 0.07$  \\ 
            &  $0.08 \pm 0.07$   &  $0.06\pm 0.02$    & $0.23 \pm 0.08$  \\
\tableline
\end{tabular}
\caption{$N_{\rm 0,HI}$ for different velocity ranges over which the \HI~emission is integrated.  The values listed in column 3 (for $\pm 0~\kms$) are the same as those listed in columns 2 and 3 of Table \ref{table2}. }\label{table3}
\end{center}
\end{table}

\begin{table}\centering
\begin{center}
\begin{tabular}{lcccc}
\multicolumn{5}{c}{Table 4: $N_{\rm 0,HI}$ for different spatial boundaries.}\\
\tableline\tableline
Cloud  &  $N_{\rm 0,HI}$ [$10^{21}~\cm$]   &  $N_{\rm 0,HI}$ [$10^{21}~\cm$]    & $N_{\rm 0,HI}$ [$10^{21}~\cm$] & $N_{\rm pixels}$  \\
       & $\sigma_{\rm HI}$ [$10^{21}~\cm$] &  $\sigma_{\rm HI}$ [$10^{21}~\cm$] &  $\sigma_{\rm HI}$ [$10^{21}~\cm$]  &    \\
       & \footnotesize{for $A_{\rm V,NICEST}\ge 0.5$ mag} &  \footnotesize{for $A_{\rm V,NICEST}\ge 1$ mag}  &  \footnotesize{for $A_{\rm V,NICEST}\ge 1.5$ mag} &\\  
\tableline
Ophiuchus   & $0.86 \pm 0.03$    &  $0.86 \pm 0.05$   & $0.84 \pm 0.11$  & 675, 341, 172 \\
            & $0.12 \pm 0.03$    &  $0.13 \pm 0.06$   & $0.13 \pm 0.12$  &   \\
Perseus     & $0.92 \pm 0.04$    &  $0.98 \pm 0.05$   & $1.01 \pm 0.09$  & 272, 92, 41  \\
            & $0.14 \pm 0.04$    &  $0.07 \pm 0.05$   & $0.06 \pm 0.10$  &  \\
Orion A     & $0.99 \pm 0.10$    &  $1.05 \pm 0.11$   & $1.04 \pm 0.15$  & 127, 99, 60   \\
            & $0.23 \pm 0.11$    &  $0.20 \pm 0.12$   & $0.22 \pm 0.18$  & \\  
Orion B     & $1.38 \pm 0.02$    &  $1.39 \pm 0.07$   & $1.40 \pm 0.10$  & 248, 132, 61 \\
            & $0.11 \pm 0.02$    &  $0.10 \pm 0.07$   & $0.12 \pm 0.10$  & \\  
California  & $1.50 \pm 0.03$    &  $1.52 \pm 0.06$   & $1.56 \pm 0.09$  & 285, 124, 59  \\
            & $0.13 \pm 0.03$    &  $0.14 \pm 0.07$   & $0.11 \pm 0.09$  & \\
MonR2       & $1.29 \pm 0.04$    &  $1.29 \pm 0.10$   & $1.36 \pm 0.07$  & 77, 33, 15\\ 
            & $0.15 \pm 0.04$    &  $0.14 \pm 0.11$   & $0.07 \pm 0.07$  & \\ 
Rosette     & $1.95 \pm 0.04$    &  $2.03 \pm 0.03$   & $2.06 \pm 0.14$  & 60, 31, 18 \\ 
            & $0.23 \pm 0.04$    &  $0.06 \pm 0.02$   & $0.04 \pm 0.15$  & \\
\tableline
\end{tabular}
\caption{$N_{\rm 0,HI}$ within different spatial boundaries, as defined by the NICEST maps.  The values listed in column 3 (for $A_{\rm V,NICEST}\ge 1$ mag) are the same as those listed in columns 2 and 3 of Table \ref{table2}.  Column 5 lists the number of pixels, $N_{\rm pixels}$, within the spatial boundaries  $A_{\rm V,NICEST}\ge 0.5$, $\ge 1$, and $\ge 1.5$ mag .  }\label{table4}
\end{center}
\end{table}

\begin{table}\centering
\begin{center}
\begin{tabular}{lcc}
\multicolumn{3}{c}{Table 5: \HI-to-\htwo~Transition.}\\
\tableline\tableline
Cloud       & $\Sigma_{\rm HI}$ at $\rhtwo=0.25$ &  $\Sigma_{0,\rm HI}$ from $N_{\rm 0,HI}$  \\  
            & [\sunits]                         &  [\sunits]   \\
\tableline
Ophiuchus   & $7.4 \pm 1.3$   &  $6.9 \pm 0.5$ \\
Perseus     & $7.0 \pm 1.4$   &  $7.9 \pm 0.3$  \\
Orion A     & $7.8 \pm 1.9$   &  $8.4 \pm 0.6$  \\  
Orion B     & $8.9 \pm 3.4$   &  $11.2 \pm 0.2$  \\  
California  & $11.4 \pm 2.4$  &  $12.2 \pm 0.5$  \\
MonR2       & $9.6 \pm 2.3$   &  $10.4 \pm 0.6$  \\  
Rosette     & $15.9 \pm 2.4$  &  $16.6 \pm 0.2$  \\ 
\tableline
\end{tabular}
\caption{Column 2 lists the \HI~surface densities at $\rhtwo=0.25\pm 0.1$. Column 3 lists the \HI~surface densities from the peaks of the \HI~PDFs, $\nhi$, using equation (\ref{eq:surface}).}\label{table5}
\end{center}
\end{table}

\end{document}